# Density fluctuations and temperature relaxation in multicomponent plasmas

B J B Crowley[1,2,3]

[1]Department of Physics, University of Oxford, Parks Road, Oxford OX1 3PU, UK
[2]AWE PLC, Reading RG7 4PR, UK

[3]Email: *basil.crowley@physics.ox.ac.uk*

Date: 02 October 2015

A general formalism for the treatment of density fluctuations in Coulomb plasmas is presented and applied to the treatment of temperature relaxation in multi-component quantum plasmas when the separate components (electrons and ions) relax to LTE much faster than the mutual equilibration rate. The underlying theory is based upon the Random Phase Approximation with static local field corrections (LFCs) to account for short-range correlations between the particles. A general formula for the electron-ion equilibration rate, similar to the one given by Daligault and Dimonte, is derived. However a different approach is advocated for the determination of the LFCs , one that allows a detailed quantal treatment of the electrons and with allowance for the possible presence of bound states. The theory presented here accommodates any number of plasma components and provides a basis for an approximate self-consistent solution of the general Coulomb many-body problem that incorporates static short-range correlations into the RPA. This has implications for improved modelling of the static and low frequency transport and equation-of-state properties of dense plasma. Limitations of the model are also discussed.

**PACS**: **52.20.-j**; 52.20Hv; 52.20Fs; **52.25.-b**; 52.25Gj, 52.25Mq; 52.25Cm

**Keywords**: Multi-component plasmas, quantum plasmas, Coulomb many-body problem, fluctuations, correlations, electron-ion equilibration, temperature relaxation, local field corrections, equation of state.

This page is intentionally left blank

# CONTENTS

# 1 INTRODUCTION

## 1.1 Temperature relaxation in multicomponent plasmas

The creation of transient hot dense plasmas for laboratory study or in ICF applications generally utilises heating mechanisms that selectively pump energy into one or other of the plasma components, nuclear ions or electrons. Electromagnetic processes supply energy mainly to electrons, while nuclear processes impart their energy primarily to nuclei. The question of how quickly the other plasma component(s) react to this and relax to a common temperature then arises. An underlying issue is that the smallness of the ratio of the electron mass to the mass of an ion inhibits the transfer of energy between these particles. In a binary collision between a fast moving electron and a quasi-stationary ion, for example, the fractional energy loss by the electron to the ion is $\sim m_e/m_i$ so that many thousands of collisions will typically be required to bring a single ion into equilibrium. The subsystems comprising these individual species can thus relax to differing temperatures on timescales that are relatively short compared to the timescales for mutual thermalization to a common temperature. The problem is particularly acute for ICF, in which a short-pulse laser is conceivably used to heat the compressed plasma to ignition temperatures following compression. However, in such a system, and for radiatively driven systems generally, the drive energy is primarily absorbed by the electrons. However the objective is to heat the atomic nuclei, in order to bring about their fusion. If the electrons are unable to transfer their excess energy to the ions fast enough, ignition will not occur before the plasma decompresses. In any case, the strong temperature dependence of the fusion burn rate means that accurate modelling of the electron-ion equilibration dynamics is essential for understanding and predicting the performance of such systems.

Conversely, in plasmas subject to radiative cooling, the ion temperature may lag behind the electron temperature if the energy exchange between ions and electrons cannot match the radiative losses. In such a system, the rate of radiative energy loss may be given by Stefan's Law as $\propto T_e^4$, while the rate of energy transfer between electrons and ions is inferred to be $\propto (m_e/m_i)(T_i - T_e)/T_e^{3/2}$, so the hotter the electrons, the harder it is for the ions to stay in equilibrium.

This problem has been the subject of extensive study over several decades. The original approach, due to Landau [1], [2] and Spitzer [3], [4] is to consider classical binary collisions between ions and electrons. However, for Coulomb interactions, this leads to a divergent collision integral, with divergences at both short and long impact parameters. The requirement to regularise the Coulomb Logarithm pervades all problems involving collisional processes in Coulomb systems, including, for example, determination of the electrical and thermal conductivities [5]. The short-range divergence, which is due to the singularity in the potential, is removed by quantum diffraction: The finite wavelength of the scattered particle means that the singularity is unresolved. At large length scales, many-body effects become important and result in the integral being effectively truncated at large distances. Unfortunately, although the reasons for the divergences may be understood, the value of the Coulomb Logarithm can, particularly for non-ideal plasmas, be quite sensitive to the modelling and, for this reason, methods using *ad hoc* cutoffs are generally unreliable.

Quantum techniques that treat the many body aspects from the outset do not generally lead to divergent integrals or require *ad hoc* corrections. However a number of differing formulae have emerged [6], [7], [8], [9], [10], [11], [12], which can lead to quite different results for given systems. Ref. [12] gives a critique of some of these formulae and places them in a hierarchy of approximations. Ref. [13] compares the results of calculations, albeit from a different subset. An outstanding problem is reconciling these different formulae and determining which, if any, are applicable in specific circumstances.

Many-body treatments of plasmas typically rely upon the Random Phase Approximation (RPA) [14],[15],[16] at some level and can thereby treat static and dynamical screening and collective effects, through the Lenard-Balescu collision integral [5],[16],[17], [18]. However, this approximation fails to treat the short-range correlations, particularly between the electrons and ions, which are those likely to have an important effect on the electron-ion relaxation. More recently, the treatment of short range correlations in the temperature relaxation of two component plasmas, at the level of a static approximation, has been achieved through the introduction of Local Field Corrections (LFCs) [12], which effectively modify the potentials so as to give rise to a more accurate pair correlation functions when applied within the RPA. This is the approach used here, and the resulting general formula for the electron-ion energy exchange rate, which is applicable to plasmas of arbitrary degeneracy, is identical to that obtained in ref. [12], apart from the method used to determine the LFCs themselves. Also the derivations used here are subtly different and appear to rely on fewer approximations. Another difference is that the present theory is generalized to any number of plasma components satisfying local continuity. A proper treatment of a D-T plasma, which is one comprising three components, without the need to average over the ion properties, therefore becomes possible. The theory presented here would apply to a multi-component plasma in which all components are weakly coupled to each other and could therefore conceivably apply to a three component plasma comprising light and heavy ion species.

The method proposed here for the determination of the LFCs involves using ion-ion and electron-ion pair correlation functions determined by separate molecular dynamics (MD) calculations or hypernetted chain (HNC) or Percus-Yevick (PY) approximations for the ion-ion correlations, combined with single-cell density functional theory (DFT) or average-atom calculations of the electronic structure. The model is also careful to treat the asymmetries introduced in some forms of these functions as a result of the assumed temperature differences between the components.

An experiment to test the various predictions was proposed at the Los Alamos National Laboratory [19] and some comparative predictions are given in ref.[13].

## 1.2 Plasma fluctuations

In many body systems in which the particles interact via long-range forces, the treatment of dynamical processes is often best achieved by considering fluctuations in the local properties of a system rather than by attempting to treat the dynamics through a series of pairwise particle interactions. For Coulomb systems, such pairwise "collisions" are always subject to modification by the other particles in the system and, in order to avoid divergences, it is typically necessary to impose *ad hoc* cutoffs, which are, especially in non-ideal plasmas, prone to error or uncertainty. The treatment of fluctuations, which include static spatial inhomogenieties as well as collective dynamics, through the use of Green functions in quantum many body theory, on the other hand, takes account of all of the particles in the system. The treatment of Coulomb interactions is rendered much more tractable by working in reciprocal ($\mathbf{q}$) space. This is because each particle effectively interacts with every other particle in the system, but with large-scale cancellations due to overall neutrality. In reciprocal space, such a convolution of a long-range potential with the charge density across the entire system is replaced with a simple product of the Fourier transform of the potential with a function embodying only the charge density fluctuations. Taking this one step further, it is the fluctuations, which are manifestations of collective behaviour, that are the fundamental entity, so, rather than considering interactions between particles, or between particles and fluctuations, the many body theory of Coulomb systems considers the interactions between fluctuations. This results in a formulation that depends upon correlation functions and associated Green functions. Whereas correlation functions are simply expectation values of products of observables at different locations in the system and/or at different times, Green functions relate the correlations between observables to the propagation of disturbances, which may originate spontaneously or as a result of external influences. Green functions are therefore related to expectation values of time-ordered operators and take account of the causal nature of dynamical processes. Thus correlation functions represent properties of the state of a many body system whereas Green functions incorporate information about how the correlations arise from kinematical and dynamical processes occurring within the system.

The Wiener-Khinchin theorem [20] states that the Fourier transform of the time autocorrelation function of a steady-state fluctuation, one whose first and second moments are constant in time, is equivalent to the power spectral density. This is a generalization of the convolution theorem, which, in the special case of observables that are square-integrable over infinite times (pulsed signals), yields a similar relation between the energy spectral density and the autocorrelation function.

Correlation functions and Green functions are intimately linked by the Fluctuation Dissipation Theorem [21], [22], [23] which directly relates the autocorrelation function representing fluctuations in some observable to the imaginary (dissipative) part of the Green function describing the response of that observable to an applied force. Statements of this theorem take many (equivalent) forms, several of which, those relating to density fluctuations, emerge in the following, but at the heart of it is a simple statement of a deep and fundamental relationship between the fluctuations in an observable and energy dissipation associated with a displacement of that observable from equilibrium by an external influence, to the extent that they can be considered to be different sides of the same coin. The fluctuation dissipation theorem is one of a number of fundamental theorems, including the aforementioned Wiener-Khinchin theorem, underpinning the statistical mechanics of non-equilibrium systems [24], [25] and thermodynamics [26] of quantum systems.

The modelling of fluctuations is a particularly appropriate means of treating the problem of thermal relaxation between different plasma components. The problem essentially concerns the interaction between two (or more) many body systems occupying the same volume of space. Fluctuating electric fields due to the charge-density fluctuations in one system will act upon the current fluctuations of another and vice versa, giving rise to exchange of energy. This is, on the whole, a more complete description of the thermalization process than that attributed to collisions between the constituent particles, especially given that, in Coulomb systems, such collisions are localized neither in space nor in time.

## 2  DENSITY FLUCTUATIONS IN MULTICOMPONENT PLASMAS – GENERAL THEORY

### 2.1  Density correlation function

We consider a homogeneous multi-component plasma comprising charge species $\{a,b,c,...\}$ in a quasi steady-state maintained at constant volume, $\mathfrak{V}$. The individual plasma components will be treated as subsystems that interact weakly with each other while being considered to be in LTE at separate temperatures $\{T_a, T_b, T_c,...\}$. However, to begin with, these are taken to be equal to a common temperature, $T$. The densities of each species are subject to spatial and temporal fluctuations $\delta n_a(\mathbf{r},t) = n_a(\mathbf{r},t) - \bar{n}_a$ where $\bar{n}_a = \frac{1}{\mathfrak{V}} \int_{\mathfrak{V}} n_a(\mathbf{r},t) d^3\mathbf{r}$ is the mean density of species $a$, which is independent of position $\mathbf{r}$ and time $t$. In terms of the density operators $\hat{n}_a(\mathbf{r},t) = e^{i\hat{H}t} \hat{n}_a(\mathbf{r},0) e^{-i\hat{H}t}$ and $\delta\hat{n}_a(\mathbf{r},t) = \hat{n}_a(\mathbf{r},t) - \bar{n}_a$, we define the dynamic structure factor, which is the space and time Fourier transform of the space-time density autocorrelation function $\langle \delta\hat{n}_a(\mathbf{r},t) \delta\hat{n}_b(\mathbf{r}',t') \rangle$, as follows

$$S_{ab}(\mathbf{q},\omega) = \frac{1}{2\pi n_e \mathfrak{V}} \iint_{\mathfrak{V}} d^3\mathbf{r}\, d^3\mathbf{r}' \int_{-\infty}^{+\infty} dt\, \langle \delta\hat{n}_a(\mathbf{r},t) \delta\hat{n}_b(\mathbf{r}',t') \rangle e^{-i\mathbf{q}\cdot(\mathbf{r}-\mathbf{r}')} e^{i\omega(t-t')} \qquad (1)$$

where the notation $\langle \hat{O} \rangle = \text{tr}(\hat{O}\hat{\rho})$ denotes the thermodynamic average or expectation value of an operator $\hat{O}$ and where $\hat{\rho} = \exp(-\hat{H}/T) / \text{tr}(\exp(-\hat{H}/T))$ is the statistical operator in the Canonical Ensemble of systems with Hamiltonian $\hat{H}$ in thermodynamic equilibrium at temperature $T$. For a homogeneous system in equilibrium (or in a steady state) the autocorrelation function of any fluctuating observable, depends only on the time displacement $t-t'$. $S_{ab}(\mathbf{q},\omega)$ given by (1) is therefore independent of time.

By inversion of the Fourier transform (1),

$$\langle \delta\hat{n}_a(\mathbf{r},t) \delta\hat{n}_b(\mathbf{r}',t') \rangle = \frac{n_e}{(2\pi)^3} \int d^3\mathbf{q} \int_{-\infty}^{+\infty} d\omega\, S_{ab}(\mathbf{q},\omega) e^{i\mathbf{q}\cdot(\mathbf{r}-\mathbf{r}')} e^{-i\omega(t-t')} \qquad (2)$$

and

$$\langle \delta\hat{n}_a(\mathbf{q},t) \delta\hat{n}_b(-\mathbf{q},t') \rangle = n_e \mathfrak{V} \int_{-\infty}^{+\infty} S_{ab}(\mathbf{q},\omega) e^{-i\omega(t-t')} d\omega$$

$$= \iint_{\mathfrak{V}} \langle \delta\hat{n}_a(\mathbf{r},t) \delta\hat{n}_b(\mathbf{r}',t') \rangle e^{-i\mathbf{q}\cdot(\mathbf{r}-\mathbf{r}')} d^3\mathbf{r}\, d^3\mathbf{r}' \qquad (3)$$

where

$$\delta\hat{n}_a(\mathbf{q},t) = \int_{\mathfrak{V}} \delta\hat{n}_a(\mathbf{r},t) e^{-i\mathbf{q}\cdot\mathbf{r}} d^3\mathbf{r} \qquad (4)$$

As shown in Appendix A.2, the matrix function $S_{ab}(\mathbf{q},\omega)$ satisfies a number of key relations:

Time reversal relation:

$$S_{ab}(\mathbf{q},\omega) = S_{ab}(-\mathbf{q},\omega) \qquad (5)$$

Hermiticity relation:

$$S_{ab}^*(\mathbf{q},\omega) = S_{ba}(\mathbf{q},\omega) \qquad (6)$$

Kubo-Martin-Schwinger (KMS) relation [16],

$$S_{ab}(\mathbf{q},-\omega) = e^{-\omega/T} S_{ba}(\mathbf{q},\omega) \tag{7}$$

### 2.2 The response function and the fluctuation dissipation theorem

The Green function giving the variation in the density in response to an external field is [23],[24]

$$\begin{aligned} \mathrm{K}_{ab}(\mathbf{q},\omega) &= -\frac{\mathrm{i}}{\mathfrak{V}}\int_{t'}^{\infty}\langle[\delta\hat{n}_a(\mathbf{q},t),\delta\hat{n}_b(-\mathbf{q},t')]\rangle e^{\mathrm{i}(t-t')(\omega+\mathrm{i}0^+)}\mathrm{d}t \\ &= -\frac{\mathrm{i}}{\mathfrak{V}}\int_{0}^{\infty}\langle\delta\hat{n}_a(\mathbf{q},t)\delta\hat{n}_b(-\mathbf{q},0) - \delta\hat{n}_b(-\mathbf{q},0)\delta\hat{n}_a(\mathbf{q},t)\rangle e^{\mathrm{i}t(\omega+\mathrm{i}0^+)}\mathrm{d}t \\ &= -\mathrm{i}n_e\int_{0}^{\infty}e^{\mathrm{i}t(\omega+\mathrm{i}0^+)}\mathrm{d}t\int_{-\infty}^{+\infty}\left(S_{ab}(\mathbf{q},\omega')e^{-\mathrm{i}\omega't} - S_{ba}(-\mathbf{q},\omega')e^{\mathrm{i}\omega't}\right)\mathrm{d}\omega' \\ &= n_e\int_{-\infty}^{+\infty}\left(\frac{S_{ab}(\mathbf{q},\omega')}{\omega-\omega'+\mathrm{i}0^+} - \frac{S_{ba}(-\mathbf{q},\omega')}{\omega+\omega'+\mathrm{i}0^+}\right)\mathrm{d}\omega' \end{aligned} \tag{8}$$

Let the matrix $\mathbf{K}$ be decomposed into its Hermitian and anti-Hermitian parts according to $\mathbf{K} = \mathbf{K'} + \mathrm{i}\mathbf{K''}$ where $\mathbf{K'}$ and $\mathbf{K''}$ are both Hermitian, where

$$\mathbf{K'} = \tfrac{1}{2}(\mathbf{K} + \mathbf{K}^{\dagger})$$
$$\mathbf{K''} = -\tfrac{1}{2}\mathrm{i}(\mathbf{K} - \mathbf{K}^{\dagger}) \tag{9}$$

so that, for example,

$$\mathrm{K}''_{ab} = -\tfrac{1}{2}\mathrm{i}\left(\mathrm{K}_{ab} - \mathrm{K}^*_{ba}\right) \tag{10}$$

Then, referring back to (8), and using the Hermitian property of $\mathbf{S}$, along with the property (5),

$$\mathrm{i}\mathrm{K}''_{ab}(\mathbf{q},\omega) = \tfrac{1}{2}n_e\int_{-\infty}^{+\infty}\left(S_{ab}(\mathbf{q},\omega')\left(\frac{1}{\omega-\omega'+\mathrm{i}0^+} - \frac{1}{\omega-\omega'-\mathrm{i}0^+}\right) - S_{ba}(\mathbf{q},\omega')\left(\frac{1}{\omega+\omega'+\mathrm{i}0^+} - \frac{1}{\omega+\omega'-\mathrm{i}0^+}\right)\right)\mathrm{d}\omega'$$
(11)

Application of the Cauchy identity,

$$\frac{1}{z \mp \mathrm{i}0^+} = \wp\left(\frac{1}{z}\right) \pm \mathrm{i}\pi\delta(z) \tag{12}$$

where $\wp$ denotes the principal value part, yields

$$K''_{ab}(\mathbf{q},\omega) = -\pi n_e \left( S_{ab}(\mathbf{q},\omega) - S_{ba}(\mathbf{q},-\omega) \right)$$

$$= -\pi n_e \left(1 - e^{-\omega/T}\right) S_{ab}(\mathbf{q},\omega) \tag{13}$$

or, in matrix form,

$$\mathbf{K}''(\mathbf{q},\omega) = -\pi n_e \left(1 - e^{-\omega/T}\right) \mathbf{S}(\mathbf{q},\omega) \tag{14}$$

which is a statement of the Fluctuation Dissipation theorem for multicomponent systems. The form of this equation is the same as for single component systems, except that, in (14), the quantities on both sides of this equation are Hermitian matrices. Moreover, equation (14) together with the KMS relation (7) imply that the real and imaginary parts of $\mathbf{K}''(\mathbf{q},\omega)$ are respectively odd and even functions of $\omega$.

### 2.3 Screening in multi-component plasmas.

For Coulomb systems, the general response Green function matrix $\mathbf{K}$ satisfies the generalized screening equation

$$\mathbf{K} = \mathbf{\Pi} + \mathbf{\Pi} v \mathbf{K} \tag{15}$$

where $\mathbf{\Pi} = \|\Pi_{ab}\|$ is the susceptibility, which is the polarization part, or irreducible part of $\mathbf{K}$, and $v = \|Q_a Q_b v(\mathbf{q})\|$ is the interaction potential matrix, the component, $Q_a Q_b v(\mathbf{q})$, of which expresses the interaction between charge species $a$ and $b$ having charges $Q_a e$ and $Q_b e$ respectively and

$$v(\mathbf{q}) = \frac{e^2}{4\pi\varepsilon_0} \int \frac{e^{-i\mathbf{q}\cdot\mathbf{r}}}{r} d^3\mathbf{r} = \frac{4\pi e^2}{4\pi\varepsilon_0 q^2} \tag{16}$$

is the Fourier transform of the Coulomb potential acting between unit charges.

Equation (15) can be expressed in the form

$$(1-\chi)\mathbf{K} = \mathbf{\Pi} \tag{17}$$

where $\chi = \mathbf{\Pi} v$ is the polarizability, which is the matrix with components given by

$$\chi_{ab} = P_a Q_b \tag{18}$$

where

$$P_a = v \sum_c \Pi_{ac} Q_c \quad (19)$$

Now

$$(\chi\chi)_{ab} = \sum_c \chi_{ac}\chi_{cb} = \mathbf{P}\cdot\mathbf{Q}\,P_a Q_b = \mathbf{P}\cdot\mathbf{Q}\,\chi_{ab} \quad (20)$$

where

$$\mathbf{P}\cdot\mathbf{Q} = v\sum_{a,b} Q_a \Pi_{ab} Q_b = \sum_a \chi_{aa} \equiv 1 - D \quad (21)$$

and where

$$D(\mathbf{q},\omega) = 1 - \mathrm{trace}(\chi(\mathbf{q},\omega)) \quad (22)$$

The matrix $\Phi = \dfrac{\chi}{1-D}$ is therefore idempotent ($\Phi^2 = \Phi$) and therefore, by lemma 2 in APPENDIX A,

$$(\mathbf{1}-\chi)^{-1} = (\mathbf{1}+(D-1)\Phi)^{-1} = \mathbf{1} - \frac{D-1}{D}\Phi = \mathbf{1} + \frac{\chi}{D} \quad (23)$$

Hence the solution of (17) is

$$\mathbf{K} = (\mathbf{1}-\chi)^{-1}\Pi = \left(\mathbf{1}+\frac{\chi}{D}\right)\Pi$$

$$= \Pi + \frac{1}{D}\Pi v \Pi \quad (24)$$

which is an exact result. Also, the idempotency property implies that

$$\chi^2 = (1-D)\chi \quad (25)$$

and hence, from (24),

$$\mathbf{K}v = \frac{\chi}{D} \quad (26)$$

in which $D(\mathbf{q},\omega)$ is recognised as being the dielectric function.

Now, according to the Fluctuation Dissipation Theorem (14), the density fluctuations, as expressed by the dynamic structure factor, depend only upon the anti-Hermitian part of $\mathbf{K}$. Accordingly it is appropriate to seek an equation for $\mathbf{K}''$, as defined by (9) or (10). To this end, let

$\mathbf{L} = \mathbf{1} + \dfrac{\chi}{D} = (\mathbf{1}-\chi)^{-1}$, and define $\mathbf{J}$ by

$$\mathbf{J} = \mathbf{L}^{-1}\mathbf{K}\mathbf{L}^{\dagger -1}$$

$$= (1-\chi)\mathbf{K}(1-\chi^{\dagger})$$

$$= \mathbf{\Pi}(1-\chi^{\dagger}) \tag{27}$$

$$= \mathbf{\Pi}(1-v\mathbf{\Pi}^{\dagger})$$

The anti-Hermitian part of $\mathbf{J}$ is then given by

$$\mathbf{J}'' = \mathbf{\Pi}'' \tag{28}$$

From the first of (27),

$$\mathbf{K} = \mathbf{L}\mathbf{J}\mathbf{L}^{\dagger} \tag{29}$$

and hence, by taking the anti-Hermitian part of both sides and applying (28),

$$\mathbf{K}'' = \mathbf{L}\mathbf{J}''\mathbf{L}^{\dagger} = \mathbf{L}\mathbf{\Pi}''\mathbf{L}^{\dagger} \tag{30}$$

which is equivalent to

$$\mathbf{K}'' = \left(1+\frac{\chi}{D}\right)\mathbf{\Pi}''\left(1+\frac{\chi^{\dagger}}{D^{*}}\right) \tag{31}$$

Equation (31) is a key relation between the dissipative (anti-Hermitian) part of the response function and its polarization part.

### 2.4　Observable correlations and the symmetrised correlation function

The non-commuting nature of the density operators $\hat{n}_a(\mathbf{r},t)$ and $\hat{n}_b(\mathbf{r}',t')$, which are themselves Hermitian, for interacting particles, means that the correlation operator, $\delta\hat{n}_a(\mathbf{r},t)\delta\hat{n}_b(\mathbf{r}',t')$ is non-Hermitian on the Fock space, and therefore does not correspond to a physical observable. One can however define the symmetrised (Hermitian) density correlation operator,

$$\hat{c}_{ab}(\mathbf{r},\mathbf{r}',t,t') = \tfrac{1}{2}\left(\delta\hat{n}_a(\mathbf{r},t)\delta\hat{n}_b(\mathbf{r}',t') + \delta\hat{n}_b(\mathbf{r}',t')\delta\hat{n}_a(\mathbf{r},t)\right)$$

$$= \hat{c}_{ab}(\mathbf{r}-\mathbf{r}',t-t') \tag{32}$$

corresponding to which is the symmetrised dynamic structure factor,

$$\widehat{S}_{ab}(\mathbf{q},\omega) = \frac{1}{2\pi n_e \mathfrak{V}} \iint_{\mathfrak{V}} d^3\mathbf{r}\, d^3\mathbf{r}' \int_{-\infty}^{+\infty} dt\, \langle \hat{c}_{ab}(\mathbf{r},\mathbf{r}',t,t') \rangle e^{-i\mathbf{q}\cdot(\mathbf{r}-\mathbf{r}')} e^{i\omega(t-t')}$$

$$= \frac{1}{2\pi n_e} \int_{\mathfrak{V}} d^3\mathbf{r} \int_{-\infty}^{+\infty} dt\, \langle \hat{c}_{ab}(\mathbf{r},t) \rangle e^{-i\mathbf{q}\cdot\mathbf{r}} e^{i\omega t} \qquad (33)$$

$$= \frac{1}{2\pi n_e} \int_{-\infty}^{+\infty} \langle \hat{c}_{ab}(\mathbf{q},t) \rangle e^{i\omega t} dt$$

where

$$\hat{c}_{ab}(\mathbf{q},t) = \int_{\mathfrak{V}} \hat{c}_{ab}(\mathbf{r},t) e^{-i\mathbf{q}\cdot\mathbf{r}} d^3\mathbf{r}$$

$$\langle \hat{c}_{ab}(\mathbf{q},t) \rangle = n_e \int_{-\infty}^{\infty} \widehat{S}_{ab}(\mathbf{q},\omega) e^{-i\omega t} d\omega \qquad (34)$$

The relationship between $S_{ab}(\mathbf{q},\omega)$ and $\widehat{S}_{ab}(\mathbf{q},\omega)$ is expressed by

$$\widehat{S}_{ab}(\mathbf{q},\omega) = \tfrac{1}{2}\left(S_{ab}(\mathbf{q},\omega) + S_{ba}(-\mathbf{q},-\omega)\right)$$

$$= \tfrac{1}{2}\left(1 + e^{-\omega/T}\right) S_{ab}(\mathbf{q},\omega) \qquad (35)$$

where use has been made of the KMS relation (7) and the time-reversal property (5). Finally, combining (35) and (14),

$$\widehat{S}(\mathbf{q},\omega) = \tfrac{1}{2}\left(1 + e^{-\omega/T}\right) S(\mathbf{q},\omega) = -\frac{1}{2\pi n_e} \coth\left(\frac{\omega}{2T}\right) K''(\mathbf{q},\omega) \qquad (36)$$

which is another statement of the Fluctuation Dissipation Theorem.

## 3 TEMPERATURE RELAXATION IN MULTICOMPONENT PLASMAS

### 3.1 General formula for energy exchange rate

We now apply the foregoing to treating the relaxation, through thermal energy exchange, of a multicomponent plasma when the components are at different temperatures $T_a \neq T_b$ etc. The relaxation time is assumed to be long compared to the timescale of fluctuations, and this will be attributed to some degree of weak coupling between the components. Such weak coupling can occur, in Coulomb systems, if one component is much more massive than another, eg, for ions and electrons, or for a mixture of heavy and light ions. The individual components are assumed to be internally in LTE with much shorter relaxation times than the mutual equilibration time(s).

Energy exchange occurs as a result of the forces acting between particles of different species and can be described in terms of the charge current fluctuations in one species interacting with the electric field fluctuations of another, whereby the net energy transfer rate $a \to b$ per unit volume is

$$\dot{\mathcal{E}}_{ab} = \frac{\partial \mathcal{E}_b}{\partial t}\bigg|_{a\to} = \frac{1}{2\mathfrak{V}} \int_{\mathfrak{V}} \left\langle \hat{\mathbf{j}}_b \cdot \hat{\mathbf{E}}_a + \hat{\mathbf{E}}_a \cdot \hat{\mathbf{j}}_b \right\rangle d^3\mathbf{r}$$

$$= -\frac{1}{2\mathfrak{V}} \int_{\mathfrak{V}} \left\langle \hat{\mathbf{j}}_b \cdot \nabla \hat{V}_a + \left(\nabla \hat{V}_a\right) \cdot \hat{\mathbf{j}}_b \right\rangle d^3\mathbf{r}$$

(37)

where the integral is taken over a volume $\mathfrak{V}$ subject to cyclic boundary conditions, which imply

$$\iint \left\langle \hat{\mathbf{j}}_b \hat{V}_a + \hat{V}_a \hat{\mathbf{j}}_b \right\rangle \cdot d\mathbf{s} = 0 \tag{38}$$

Then, according to Gauss' theorem,

$$\int_{\mathfrak{V}} \nabla \cdot \left\langle \hat{\mathbf{j}}_b \hat{V}_a + \hat{V}_a \hat{\mathbf{j}}_b \right\rangle d^3\mathbf{r} \equiv \int_{\mathfrak{V}} \left\langle \nabla \cdot \hat{\mathbf{j}}_b \hat{V}_a + \hat{V}_a \nabla \cdot \hat{\mathbf{j}}_b \right\rangle d^3\mathbf{r} + \int_{\mathfrak{V}} \left\langle \hat{\mathbf{j}}_b \cdot \nabla \hat{V}_a + \nabla \hat{V}_a \cdot \hat{\mathbf{j}}_b \right\rangle d^3\mathbf{r} = 0 \tag{39}$$

whereupon, combining (37) and (39) yields

$$\dot{\mathcal{E}}_{ab} = \frac{1}{2\mathfrak{V}} \int_{\mathfrak{V}} \left\langle \nabla \cdot \hat{\mathbf{j}}_b \hat{V}_a + \hat{V}_a \nabla \cdot \hat{\mathbf{j}}_b \right\rangle d^3\mathbf{r} \tag{40}$$

where the operator derivatives are defined so that

$$\frac{\partial}{\partial x} \langle \hat{O} \rangle = \left\langle \frac{\partial \hat{O}}{\partial x} \right\rangle \tag{41}$$

Using the charge continuity equation for species $b$,

$$\nabla \cdot \hat{\mathbf{j}}_b + Q_b \frac{\partial \hat{n}_b}{\partial t} = 0 \tag{42}$$

(40) becomes

$$\dot{\mathcal{E}}_{ab} = -\frac{Q_b}{2\mathfrak{V}} \int_{\mathfrak{V}} \left\langle \frac{\partial \hat{n}_b}{\partial t} \hat{V}_a + \hat{V}_a \frac{\partial \hat{n}_b}{\partial t} \right\rangle d^3\mathbf{r}$$

$$= -\frac{Q_b}{2\mathfrak{V}} \int_{\mathfrak{V}} \frac{\partial}{\partial t} \left\langle \hat{n}_b(\mathbf{r},t) \hat{V}_a(\mathbf{r},0) + \hat{V}_a(\mathbf{r},0) \hat{n}_b(\mathbf{r},t) \right\rangle \bigg|_{t=0} d^3\mathbf{r} \tag{43}$$

$$= -\frac{Q_b}{2(2\pi)^3 \mathfrak{V}} \int \frac{\partial}{\partial t} \left\langle \hat{n}_b(\mathbf{q},t) \hat{V}_a(-\mathbf{q},0) + \hat{V}_a(-\mathbf{q},0) \hat{n}_b(\mathbf{q},t) \right\rangle \bigg|_{t=0} d^3\mathbf{q}$$

where $\hat{V}_a(\mathbf{q},t) = Q_a \hat{n}_a(\mathbf{q},t) v(\mathbf{q})$ is the Fourier transform of the potential due to species $a$. Substituting accordingly, and making use of (32) and (34)

$$\begin{aligned}\dot{\mathcal{E}}_{ab} &= -\frac{Q_a Q_b}{2(2\pi)^3 \mathfrak{V}} \int \frac{\partial}{\partial t}\langle \hat{n}_b(\mathbf{q},t)\hat{n}_a(-\mathbf{q},0) + \hat{n}_a(-\mathbf{q},0)\hat{n}_b(\mathbf{q},t)\rangle\Big|_{t=0} v(\mathbf{q}) \mathrm{d}^3\mathbf{q} \\ &= -\frac{Q_a Q_b}{(2\pi)^3} \int \frac{\partial}{\partial t}\langle \hat{c}_{ab}(\mathbf{q},t)\rangle\Big|_{t=0} v(\mathbf{q}) \mathrm{d}^3\mathbf{q} \\ &= \frac{\mathrm{i} Q_a Q_b n_e}{(2\pi)^3} \iint_{-\infty}^{+\infty} \omega \widehat{S}_{ab}(\mathbf{q},\omega) v(\mathbf{q}) \mathrm{d}\omega \mathrm{d}^3\mathbf{q} \\ &= \frac{Q_a Q_b}{(2\pi)^4} \iint_{-\infty}^{+\infty} Y_{ab}(\mathbf{q},\omega) v(\mathbf{q}) \mathrm{d}\omega \mathrm{d}^3\mathbf{q}\end{aligned}$$

(44)

which expresses the energy exchange rate in terms of an integral, over frequency and wavenumber, of the thermal coupling matrix,

$$\mathbf{Y}(\mathbf{q},\omega) = 2\pi \mathrm{i} n_e \omega \widehat{\mathbf{S}}(\mathbf{q},\omega) \tag{45}$$

Note that the energy exchange rate, (44), depends only on the part of $Y_{ab}(\mathbf{q},\omega)$ that is an even function of $\omega$.

We start by considering the system in complete thermal equilibrium (no temperature separation) so that $T = T_a = T_b$ ... etc. In this case, combining (45) with (31) and (36),

$$\mathbf{Y} = \left(1 + \frac{\boldsymbol{\chi}}{D}\right)\mathbf{X}\left(1 + \frac{\boldsymbol{\chi}^\dagger}{D^*}\right) \tag{46}$$

where

$$\mathbf{X}(\mathbf{q},\omega) = -\mathrm{i}\omega\coth\left(\frac{\omega}{2T}\right)\boldsymbol{\Pi}''(\mathbf{q},\omega) \tag{47}$$

### 3.2 Weak coupling approximation with local field corrections

We now make a weak coupling approximation for the coupling between different species, or species groups, by neglecting the off-diagonal elements of the susceptibility, ie $\Pi_{ab} = 0$ for $a \neq b$. This is consistent with the RPA and the Lindhard formula, which apply in situations of weak coupling generally. However, at this stage, no approximations are necessarily implied in respect of the diagonal elements, $\Pi_{aa}$, which in principle, could be provided exactly for each subsystem, considered in isolation, in terms of the separate temperatures $T_a$. By this means, there is the possibility of being able to model the temperature separation between the plasma components, without having to make a weak-coupling approximation for those individual components themselves.

In practice however, one generally has to resort to using the RPA Lindhard formula for $\Pi_{aa}$. However such approximations disregard the correlations between the various components at short distances, as well as the effect of the mixing of the different species on the correlations between particles of the same species. These potentially important missing correlation effects can be incorporated through the use of static local field corrections (LFCs) [27], according to which the static correlations are represented through a systematic replacement of $v(\mathbf{q})$ with an effective potential $v^{\text{eff}}(\mathbf{q}) = \left\| v_{ab}^{\text{eff}}(\mathbf{q}) \right\|$, which is expressed by

$$v_{ab}^{\text{eff}}(\mathbf{q}) = v_{ba}^{\text{eff}}(\mathbf{q}) = Q_a Q_a \Lambda_{ab}(\mathbf{q}) v(\mathbf{q}) \tag{48}$$

By this means, neglected off-diagonal terms, as well as approximations necessarily used in the calculation of the functions $\Pi_{aa}(\mathbf{q},\omega)$ can, at least in a static approximation, be accounted for. Because they represent a static potential, the functions $\Lambda_{ab}(\mathbf{q})$ are necessarily real and symmetric ($\Lambda_{ab} = \Lambda_{ba}$).

In principle, LFCs can be used to represent dynamical correlations as well, but the theory is then far less tractable, since the exact determination of the dynamical LFCs would be tantamount to finding an exact solution to the many body problem, making other approximations inevitable. It seems reasonable to assume that the short-range spatial correlations are the dominant correction to the RPA in hot fully-ionised systems. More generally, dynamical corrections may be required to account for processes where the components can form transient bound or resonant states through recombination and ionization and formation of transient molecules, and for the effect of localized radiative processes on electron-ion coupling generally.

In this *static local field approximation*, the Green function that gives the response to the effective field is given by, making use of (15) and (24),

$$K_{ab}(\mathbf{q},\omega) = \Pi^0_{aa}(\mathbf{q},\omega)\delta_{ab} + \Pi^0_{aa}(\mathbf{q},\omega)\sum_c v^{\text{eff}}_{ac}(\mathbf{q})K_{cb}(\mathbf{q},\omega)$$

(49)

$$= \Pi^0_{aa}(\mathbf{q},\omega)\delta_{ab} + \frac{Q_a Q_b}{D(\mathbf{q},\omega)}\Pi^0_{aa}(\mathbf{q},\omega)\Lambda_{ab}(\mathbf{q})v(\mathbf{q})\Pi^0_{bb}(\mathbf{q},\omega)$$

in which the functions $\Pi^0_{aa}$ are just the polarization parts of the one-component RPA response functions

$$K^0_{aa}(\mathbf{q},\omega) = \Pi^0_{aa}(\mathbf{q},\omega) + Q_a^2 \Pi^0_{aa}(\mathbf{q},\omega)v(\mathbf{q})K^0_{aa}(\mathbf{q},\omega) \qquad (50)$$

and are provided by the Lindhard formula. The effective polarizability $\chi$ and the effective dielectric function $D(\mathbf{q},\omega)$ are now redefined, with respect to the effective potentials and the Lindhard functions, by

$$\chi_{ab}(\mathbf{q},\omega) = Q_a Q_b \Pi^0_{aa}(\mathbf{q},\omega)\Lambda_{ab}(\mathbf{q})v(\mathbf{q}) \qquad (51)$$

$$D(\mathbf{q},\omega) = 1 - \text{trace}(\chi(\mathbf{q},\omega)) \qquad (52)$$

where, for consistency, the LFCs are defined to have the property

$$\sum_c Q_c^2 \Pi^0_{cc}(\Lambda_{ac}\Lambda_{cb} - \Lambda_{ab}\Lambda_{cc}) = 0, \qquad \forall\, a,b \qquad (53)$$

necessary to preserve the essential idempotency property (23) of the polarizability matrix (51), which, for example, maintains the solution for the response function expressed by (49). For a two component plasma, (53) is equivalent to $|\Lambda| = 0$, ie

$$\Lambda_{ab}\Lambda_{ba} = \Lambda_{aa}\Lambda_{bb}, \qquad \forall\, a,b \qquad (54)$$

while, for a plasma comprising 3 or more charged components,

$$\Lambda_{ab}\Lambda_{bc}\Lambda_{ca} = \Lambda_{aa}\Lambda_{bb}\Lambda_{cc}, \qquad \forall\, a,b,c \qquad (55)$$

which incorporates (54). These, combined with basic symmetry property of the LFC matrix, implies that the multicomponent LFCs are separable, ie $\Lambda_{ab}(\mathbf{q}) = \lambda_a(\mathbf{q})\lambda_b(\mathbf{q})$ which, note, does not imply that the functions $\lambda_a(\mathbf{q})$ are independent of the presence or properties of other components.

The real and imaginary parts of the functions $\Pi^0_{aa}(\mathbf{q},\omega)$ and $\chi_{ab}(\mathbf{q},\omega)$, thus defined, are respectively even and odd functions of $\omega$, ie,

$$\chi^*_{ab}(\mathbf{q},\omega) = \chi_{ab}(\mathbf{q},-\omega)$$

(56)

$$\Pi^{0*}_{aa}(\mathbf{q},\omega) = \Pi^0_{aa}(\mathbf{q},-\omega)$$

The components of the matrix $\mathbf{X}$ defined by (47) are given by

$$X_{ab}(\mathbf{q},\omega) = -i\omega\coth\left(\frac{\omega}{2T_a}\right)\Pi''_{aa}(\mathbf{q},\omega)\delta_{ab} \tag{57}$$

in which $\Pi''_{aa}(\mathbf{q},\omega) = \mathrm{Im}\,\Pi^0_{aa}(\mathbf{q},\omega)$, and which are therefore imaginary and odd functions of $\omega$. The elements of the matrix $\mathbf{X}$ therefore likewise satisfy a relation of the form of (56).

Expanding equation (46) leads to

$$Y_{ab} = X_{aa}\delta_{ab} + \frac{\chi_{ab}}{D}X_{bb} + \frac{\chi^*_{ba}}{D^*}X_{aa} + \frac{1}{|D|^2}\sum_c \chi_{ac}X_{cc}\chi^*_{bc}$$

$$= \frac{1}{|D|^2}\left(|D|^2\delta_{ab}X_{aa} + D^*\chi_{ab}X_{bb} + D\chi^*_{ba}X_{aa} + \sum_c \chi_{ac}\chi^*_{bc}X_{cc}\right) \tag{58}$$

which yields, for the diagonal elements of $\mathbf{Y}$,

$$Y_{aa} = \frac{1}{|D|^2}\left(|D+\chi_{aa}|^2 X_{aa} + \sum_{c\neq a}|\chi_{ac}|^2 X_{cc}\right) \tag{59}$$

and, for the off-diagonal terms, $a \neq b$,

$$Y_{ab} = \frac{1}{|D|^2}\left((D^* + \chi^*_{bb})\chi_{ab}X_{bb} + (D+\chi_{aa})\chi^*_{ba}X_{aa} + \sum_{c\neq a,b}\chi_{ac}\chi^*_{bc}X_{cc}\right) \tag{60}$$

The diagonal elements (59) are all odd functions of $\omega$, implying that $\dot{\varepsilon}_{aa} \equiv 0$, in accordance with (44). Since the elements of $\mathbf{Y}$ are generally representable as sums of products of functions with the property expressed by (56), with real coefficients, it follows that $\mathbf{Y}$ likewise has this property, ie $\mathbf{Y}^*(\mathbf{q},\omega) = \mathbf{Y}(\mathbf{q},-\omega)$, and hence the even part of $\mathbf{Y}$ is equal to $\mathrm{Re}\,\mathbf{Y}$. The integral (44) is therefore equivalent to

$$\dot{\mathcal{E}}_{ab} = \frac{Q_a Q_b}{(2\pi)^4}\iint_{-\infty}^{+\infty}\mathrm{Re}(Y_{ab}(\mathbf{q},\omega))v(\mathbf{q})\,\mathrm{d}\omega\,\mathrm{d}^3\mathbf{q} \tag{61}$$

Which vanishes for $a = b$, and in which, for $a \neq b$, the elements $Y_{ab}$ of the thermal coupling matrix are given by (60).

### 3.3 Two component plasma

In the case of a system comprising just two components, $a$ and $b$, equation (60) yields, making use of (51) and (57),

$$\operatorname{Re} Y_{ab} = \frac{1}{|D|^2} \operatorname{Re}\left( \left(D^* + \chi_{bb}^*\right) \chi_{ab} X_{bb} + \left(D + \chi_{aa}\right) \chi_{ba}^* X_{aa} \right)$$

$$= \frac{1}{|D|^2} \operatorname{Re}\left( \left(1 - \chi_{aa}^*\right) \chi_{ab} X_{bb} + \left(1 - \chi_{bb}\right) \chi_{ba}^* X_{aa} \right)$$

(62)

$$= \frac{i}{|D|^2} \left( \chi_{ab}'' X_{bb} - \chi_{ba}'' X_{aa} \right)$$

$$= \frac{Q_a Q_b}{|D(\mathbf{q},\omega)|^2} v(\mathbf{q}) \Lambda_{ab}(\mathbf{q}) \Pi_{aa}''(\mathbf{q},\omega) \Pi_{bb}''(\mathbf{q},\omega) \omega \left( \coth\left(\frac{\omega}{2T_b}\right) - \coth\left(\frac{\omega}{2T_a}\right) \right)$$

substitution of which into (61) yields

$$\dot{\mathcal{E}}_{ab} = \frac{(Q_a Q_b)^2}{(2\pi)^4} \iint_{-\infty}^{+\infty} \frac{|v(\mathbf{q})|^2}{|D(\mathbf{q},\omega)|^2} \Lambda_{ab}(\mathbf{q}) \Pi_{aa}''(\mathbf{q},\omega) \Pi_{bb}''(\mathbf{q},\omega) \left( \coth\left(\frac{\omega}{2T_b}\right) - \coth\left(\frac{\omega}{2T_a}\right) \right) \omega \, \mathrm{d}\omega \mathrm{d}^3 \mathbf{q} \quad (63)$$

where the effective dielectric function $D(\mathbf{q},\omega)$ can be expressed in terms of the single-component RPA dielectric functions,

$$\varepsilon_a^0(\mathbf{q},\omega) = 1 + Q_a^2 v(\mathbf{q}) \Pi_{aa}^0(\mathbf{q},\omega) \tag{64}$$

by

$$D(\mathbf{q},\omega) = 1 + \Lambda_{aa}(\mathbf{q})\left(\varepsilon_a^0(\mathbf{q},\omega) - 1\right) + \Lambda_{bb}(\mathbf{q})\left(\varepsilon_b^0(\mathbf{q},\omega) - 1\right) \tag{65}$$

Now, the functions $\Pi_{aa}''(\mathbf{q},\omega)$ are just those which apply to systems comprising solely the single species $a$, in the absence of other charge components. Applying the Fluctuation dissipation Theorem to such a system gives,

$$\Pi_{aa}''(\mathbf{q},\omega) = -\pi n_a \left(1 - e^{-\omega/T_a}\right) S_{aa}^0(\mathbf{q},\omega) \tag{66}$$

where $S_{aa}^0(\mathbf{q},\omega)$ is the non-interacting dynamic structure factor for species $a$, in terms of which,

$$\mathrm{K}_{aa}^{0\,\prime\prime}(\mathbf{q},\omega) = -\pi n_a \left(1 - \mathrm{e}^{-\omega/T_a}\right) \frac{S_{aa}^0(\mathbf{q},\omega)}{\left|\varepsilon_a^0(\mathbf{q},\omega)\right|^2} \quad (67)$$

which is the RPA response function for species $a$, calculated ignoring the presence of other species.

Substituting for $\Pi''_{aa}(\mathbf{q},\omega)$ and $\Pi''_{bb}(\mathbf{q},\omega)$ according to (66) in (63) yields

$$\dot{\mathcal{E}}_{ab} = \frac{(Q_a Q_b)^2 n_a n_b}{8\pi^2} \iint_{-\infty}^{+\infty} \frac{|v(\mathbf{q})|^2}{|D(\mathbf{q},\omega)|^2} \Lambda_{ab}(\mathbf{q}) S_{aa}^0(\mathbf{q},\omega) S_{bb}^0(\mathbf{q},\omega) \left(\mathrm{e}^{-\omega/T_b} - \mathrm{e}^{-\omega/T_a}\right) \omega \, \mathrm{d}\omega \, \mathrm{d}^3\mathbf{q} \quad (68)$$

which is a form of the general result for a two-component plasma. The result can also be represented in terms of the functions

$$\tilde{\mathrm{K}}_a(\mathbf{q},\omega) \equiv \frac{\Pi_{aa}^0(\mathbf{q},\omega)}{1 - \chi_{aa}(\mathbf{q},\omega)} \quad (69)$$

which are the one-component response functions including the local field corrections. The imaginary part of (69) is given by

$$\tilde{\mathrm{K}}_a''(\mathbf{q},\omega) = \frac{\Pi_{aa}''(\mathbf{q},\omega)}{\left|1 - \chi_{aa}(\mathbf{q},\omega)\right|^2} \quad (70)$$

using the fact that the LFCs are real. Using the property $\Lambda_{aa}\Lambda_{bb} = \Lambda_{ab}\Lambda_{ba}$, the effective dielectric function (65) can be expressed as follows

$$D(\mathbf{q},\omega) = (1 - \chi_{aa})(1 - \chi_{bb}) - \chi_{ab}\chi_{ba}$$

$$= (1 - \chi_{aa})(1 - \chi_{bb})\left(1 - |v_{ab}(\mathbf{q})|^2 \Lambda_{ab}(\mathbf{q}) \Lambda_{ba}(\mathbf{q}) \tilde{\mathrm{K}}_a(\mathbf{q},\omega) \tilde{\mathrm{K}}_b(\mathbf{q},\omega)\right) \quad (71)$$

where $v_{ab}(\mathbf{q}) = Q_a Q_b v(\mathbf{q})$.

Combining (70) and (71) with (63) yields another form of the main result

$$\dot{\mathcal{E}}_{ab} = \int \frac{\mathrm{d}^3\mathbf{q}}{(2\pi)^3} \int_{-\infty}^{+\infty} \frac{\mathrm{d}\omega}{2\pi} \omega \left(\coth\left(\frac{\omega}{2T_b}\right) - \coth\left(\frac{\omega}{2T_a}\right)\right) \frac{|v_{ab}(\mathbf{q})|^2 \Lambda_{ab}(\mathbf{q}) \tilde{\mathrm{K}}_a''(\mathbf{q},\omega) \tilde{\mathrm{K}}_b''(\mathbf{q},\omega)}{\left|1 - |v_{ab}(\mathbf{q})|^2 \Lambda_{ab}(\mathbf{q}) \Lambda_{ba}(\mathbf{q}) \tilde{\mathrm{K}}_a(\mathbf{q},\omega) \tilde{\mathrm{K}}_b(\mathbf{q},\omega)\right|^2} \quad (72)$$

which, upon noting that $\left(\coth\left(\frac{\omega}{2T_b}\right) - \coth\left(\frac{\omega}{2T_a}\right)\right)$ is the odd part of $2\left(\left(1-e^{-\omega/T_b}\right)^{-1} - \left(1-e^{-\omega/T_a}\right)^{-1}\right)$, is identical to the general formula derived by Daligault and Dimonte [12]. Note also that, since $\Lambda_{ab} = \Lambda_{ba}$, it follows that $\dot{\mathcal{E}}_{ab} \equiv -\dot{\mathcal{E}}_{ba}$, which is an essential requirement for a system of two components.

### 3.4 Electron–ion energy exchange

Equations (68) and (72) are among the most general expressions currently available for the energy transfer rate between the components of a two-component plasma treated as sub-systems that are internally in LTE at different temperatures $T_a$ and $T_b$. We now consider further approximations to this expression by considering it in the context of a specific typical application, namely the transfer between electrons ($a \to e$) and ions ($b \to i$). In this notation, equation (68) becomes

$$\dot{\mathcal{E}}_{ei} = \frac{Z^2 n_e n_i}{8\pi^2} \int\int_{-\infty}^{+\infty} \frac{|v(\mathbf{q})|^2}{|D(\mathbf{q},\omega)|^2} \Lambda_{ei}(\mathbf{q}) S^0_{ee}(\mathbf{q},\omega) S^0_{ii}(\mathbf{q},\omega)\left(e^{-\omega/T_i} - e^{-\omega/T_e}\right) \omega\, d\omega\, d^3\mathbf{q} \quad (73)$$

in which

$$D(\mathbf{q},\omega) = 1 + \Lambda_{ee}(\mathbf{q})\left(\varepsilon_e^0(\mathbf{q},\omega)-1\right) + \Lambda_{ii}(\mathbf{q})\left(\varepsilon_i^0(\mathbf{q},\omega)-1\right) \quad (74)$$

is the effective dielectric function (in the context of RPA) and where $Z$ is the ion charge. For this system, the weak coupling between the electrons and ions is a consequence of the smallness of the electron-ion mass ratio, $m_e/m_i$. In binary collisions between electrons and relatively slow-moving ions, for example, the mean energy transfer per collision is $\mathcal{O}(m_e/m_i)$ so that the timescales for equilibration of the electron and ion subsystems will generally be much shorter than the mutual equilibration time.

The property $m_e \ll m_i$ also has implications for approximating (73), the integral in which possesses certain similarities to the Lenard-Balescu integral for the conductivity given by equation (34) in ref. [5]. Following reference [5], we write

$$S_{ee}(\mathbf{q},\omega) = \frac{S^0_{ee}(\mathbf{q},\omega)}{|\varepsilon_e(\mathbf{q},\omega)|^2}$$

$$\tilde{S}_{ii}(\mathbf{q},\omega) = S^0_{ii}(\mathbf{q},\omega) \frac{|\varepsilon_e(\mathbf{q},\omega)|^2}{|D(\mathbf{q},\omega)|^2}$$

(75)

in which $\varepsilon_e(\mathbf{q},\omega) = 1 + \Lambda_{ee}(\mathbf{q})(\varepsilon_e^0(\mathbf{q},\omega) - 1)$ and in terms of which

$$\dot{\mathcal{E}}_{ei} = \frac{Z^2 n_e n_i}{8\pi^2} \int\int_{-\infty}^{+\infty} |v(\mathbf{q})|^2 \Lambda_{ei}(\mathbf{q}) S_{ee}(\mathbf{q},\omega) \tilde{S}_{ii}(\mathbf{q},\omega) \left(e^{-\omega/T_i} - e^{-\omega/T_e}\right) \omega \, d\omega \, d^3\mathbf{q} \tag{76}$$

Now we make the same general approximations as in ref [5], namely, since the integral over $\omega$ is dominated by low frequencies in the vicinity of the ion plasma frequency, $\Omega_i$, at frequencies very much above this, $\tilde{S}_{ii}(\mathbf{q},\omega)$ becomes negligibly small. If it is assumed that the temperatures $T_e$ and $T_i$ are such that $T_i, T_e \gg \Omega_i$, then we can make the approximation

$$e^{-\omega/T_i} - e^{-\omega/T_e} \simeq \frac{\omega}{T_e T_i}(T_e - T_i) \tag{77}$$

for $\omega \lesssim \Omega_i$, which leads to

$$\dot{\mathcal{E}}_{ei} \simeq \xi_{ei}(T_e - T_i) \tag{78}$$

where the electron-ion exchange coefficient $\xi_{ei} = \xi_{ie}$ is given by

$$\xi_{ei} = \frac{Z^2 n_e n_i}{8\pi^2 T_e T_i} \int\int_{-\infty}^{+\infty} |v(\mathbf{q})|^2 \Lambda_{ei}(\mathbf{q}) S_{ee}(\mathbf{q},\omega) \tilde{S}_{ii}(\mathbf{q},\omega) \, \omega^2 \, d\omega \, d^3\mathbf{q} \tag{79}$$

Note that the linear form of (78) does not, in this approximation, depend upon the temperature difference $|T_e - T_i|$ being small, only that both temperatures are much greater than the ion plasma frequency. If one or other of the temperatures is smaller than this, then a linear formula still results if it can be assumed that $|T_e - T_i| \ll 2T_e T_i / \Omega_i$. However it will be seen that this introduces an additional factor of $e^{-\omega/T}$, where $1/T = \frac{1}{2}(1/T_e + 1/T_i)$, into the integrand of (79). However, for most envisaged applications of this theory, the high temperature approximation is expected to be appropriate.

Focussing on the integral $\int_{-\infty}^{+\infty} S_{ee}(\mathbf{q},\omega) \tilde{S}_{ii}(\mathbf{q},\omega) \, \omega^2 \, d\omega$ contained in (79) and, following the argument given in ref. [5], we make the approximation,

$$\int_{-\infty}^{+\infty} S_{ee}(\mathbf{q},\omega) \tilde{S}_{ii}(\mathbf{q},\omega) \, \omega^2 \, d\omega \simeq S_{ee}(\mathbf{q},0) \int_{-\infty}^{+\infty} \tilde{S}_{ii}(\mathbf{q},\omega) \, \omega^2 \, d\omega \tag{80}$$

Evaluating $S_{ee}(\mathbf{q},0)$ using (75) gives [5]

$$S_{ee}(\mathbf{q},0) = \frac{1}{|\varepsilon_e(\mathbf{q},0)|^2} \frac{m_e^2 T_e}{2\pi^2 n_e q} p\left(\tfrac{1}{2}q\right) \tag{81}$$

where $p(k) = \left[1 + \exp(k^2/2m_e T_e - \eta_e)\right]^{-1}$ is the free-electron Fermi-distribution. The remaining integral over the ion structure factor can be evaluated in the Generalized Plasmon Pole Approximation [28] to give

$$\int_{-\infty}^{+\infty} \tilde{S}_{ii}(\mathbf{q}, \omega) \omega^2 \, d\omega = \frac{q^2 T_i}{m_i} (1 + \Delta(q)) \tag{82}$$

where

$$\Delta(q) = F(q) \frac{q^2}{4m_i T_i} + (1 - F(q)) \left( \frac{\Omega_q}{2T_i} \coth\left(\frac{\Omega_q}{2T_i}\right) - 1 \right) \tag{83}$$

comprises the quantal corrections, including those associated with ion plasma modes, in which

$$F(q) = \frac{\varepsilon_e(q,0)}{D(q,0)} \left( 1 - \frac{(\varepsilon_i(q,0) - 1)\Omega_i^2}{(\varepsilon_i(q,0) - 1)\Omega_q^2 - \Omega_i^2 \varepsilon_e(q,0)} \right) \tag{84}$$

$\Omega_q$ is the frequency of the ion-acoustic plasma mode $\mathbf{q}$ ($D(\mathbf{q}, \Omega_q) = 0$, $\Omega_{q=0} = \Omega_i$), $\varepsilon_i(\mathbf{q}, \omega) = 1 + D(\mathbf{q}, \omega) - \varepsilon_e(\mathbf{q}, \omega) = 1 + \Lambda_{ii}(\mathbf{q})(\varepsilon_i^0(\mathbf{q}, \omega) - 1)$ and $D(\mathbf{q}, \omega)$ is given by (74).

Finally, combining equations (79) - (82) and making use of (16),

$$\xi_{ei} = \left(\frac{Ze^2}{4\pi\varepsilon_0}\right)^2 \frac{4n_i m_e^2}{\pi m_i} (1 + e^{-\eta_e})^{-1} \ln \Lambda \tag{85}$$

$$= \frac{8\sqrt{\pi}}{T_e} \frac{m_e}{m_i} \frac{n_e^* n_i}{(2m_e T_e)^{1/2}} \left\{ \frac{\sqrt{\pi}}{2(1 + e^{-\eta_e}) I_{1/2}(\eta_e)} \right\} \left(\frac{Ze^2}{4\pi\varepsilon_0}\right)^2 \ln \Lambda$$

where $I_j(x)$ denotes the Fermi function,

$$I_j(x) = \int_0^\infty \frac{y^j}{1 + \exp(y - x)} \, dy \tag{86}$$

and

$$\ln \Lambda = (1 + e^{-\eta_e}) \int_0^\infty \frac{\Lambda_{ei}(q)}{|\varepsilon_e(q,0)|^2} (1 + \Delta(q)) p(\tfrac{1}{2}q) \frac{dq}{q}$$

$$= \int_0^\infty \frac{\Lambda_{ei}(q)}{|\varepsilon_e(q,0)|^2} (1 + \Delta(q)) \frac{1 + e^{-\eta_e}}{1 + e^{q^2/8m_e T_e - \eta_e}} \frac{dq}{q} \tag{87}$$

is the Coulomb Logarithm [5], and where the effective free-electron density (see section 4.2) $n_e^*$ is related to the degeneracy parameter $\eta_e$ by the Thomas-Fermi relation,

$$n_e^* = \frac{(2m_e T_e)^{3/2}}{2\pi^2} I_{1/2}(\eta_e) \tag{88}$$

The general form of (85) provides a definitive formula for the Coulomb Logarithm, one that incorporates, through its various factors, short-range correlations through local field corrections ($\Lambda_{ab}(\mathbf{q})$) plasma quantum effects ($\Delta(q)$) plasma screening ($|\varepsilon_e(q,0)|^{-2}$) and quantum diffraction ($(1+e^{-\eta_e})p(\tfrac{1}{2}q)$) in a single general *ab initio* formula. The presence and potential importance of plasma quantum effects has been pointed out by Gregori and Gericke [29], who consider the problem in the context of the standard plasmon-pole approximation and who also suggest some possible appropriate formulae for the ion-acoustic dispersion relation.

In the case of a non-degenerate (Boltzmann) plasma, equations (85) and (87) reduce to

$$\xi_{ei} = \frac{8\sqrt{\pi}}{T_e} \frac{m_e}{m_i} \frac{n_e^* n_i}{(2m_e T_e)^{1/2}} \left(\frac{Ze^2}{4\pi\varepsilon_0}\right)^2 \ln\Lambda \tag{89}$$

$$\ln\Lambda = \int_0^\infty \frac{\Lambda_{ei}(q)}{|\varepsilon_e(q,0)|^2} (1+\Delta(q)) e^{-q^2/8m_e T_e} \frac{dq}{q} \tag{90}$$

while, in the limit when the electrons are extremely degenerate, $T_e \ll T_F = k_F^2/2m_e$,

$$\xi_{ei} = \frac{6\pi}{T_F} \frac{m_e}{m_i} \frac{n_e^* n_i}{(2m_e T_F)^{1/2}} \left(\frac{Ze^2}{4\pi\varepsilon_0}\right)^2 \ln\Lambda \tag{91}$$

$$\ln\Lambda = \int_0^{2k_F} \frac{\Lambda_{ei}(q)}{|\varepsilon_e(q,0)|^2} (1+\Delta(q)) \frac{dq}{q} \tag{92}$$

where $k_F$ is the Fermi momentum.

### 3.5 Multi-component plasmas with more than two components

The treatment of plasmas with more than two components is less straightforward, but nevertheless interesting, because of the additional coupling terms in equation (60). We proceed by first expressing the result (63), for the case of a plasma with just two components, in the following form:

$$\dot{\mathcal{E}}_{ab} = \frac{1}{(2\pi)^4} \int \int_{-\infty}^{+\infty} F_{ab}(\mathbf{q},\omega) \Delta_{ba}(\omega) \, d\omega d^3\mathbf{q} \tag{93}$$

where

$$F_{ab}(\mathbf{q},\omega) = F_{ba}(\mathbf{q},\omega) = \frac{|v_{ab}(\mathbf{q})|^2}{|D(\mathbf{q},\omega)|^2} \Lambda_{ab}(\mathbf{q}) \Pi''_{aa}(\mathbf{q},\omega) \Pi''_{bb}(\mathbf{q},\omega) \tag{94}$$

$$\Delta_{ba}(\omega) = f_b(\omega) - f_a(\omega) \tag{95}$$

and

$$f_a(\omega) = \omega \coth\left(\frac{\omega}{2T_a}\right) \tag{96}$$

and, in terms of which it is useful to define

$$\phi_{ab}(\mathbf{q},\omega) \equiv v_{ab}(\mathbf{q}) \operatorname{Re}(Y_{ab}(\mathbf{q},\omega)) = -F_{ab}(\mathbf{q},\omega) \Delta_{ab}(\omega) \tag{97}$$

In the case of a plasma comprising an arbitrary number of components, equation (60) may be written

$$Y_{ab} = \frac{1}{|D|^2} \left( (1-\chi^*_{aa}) \chi_{ab} X_{bb} + (1-\chi_{bb}) \chi^*_{ba} X_{aa} + \sum_{c \neq a,b} (\chi_{ac} \chi^*_{bc} X_{cc} - \chi^*_{cc} \chi_{ab} X_{bb} - \chi_{cc} \chi^*_{ba} X_{aa}) \right) \tag{98}$$

in which $\chi_{ab} = \Lambda_{ab}(\mathbf{q}) v_{ab}(\mathbf{q}) \Pi_{aa}(\mathbf{q},\omega)$ are the components of the polarizability matrix (51) and $X_{aa}(\mathbf{q},\omega) = -i f_a(\omega) \Pi''_{aa}(\mathbf{q},\omega)$ are the diagonal elements of the matrix (47). Substituting accordingly and taking the real part, making use of (55) and the symmetry of the LFCs, yields

$$\phi_{ab} \equiv v_{ab}(\mathbf{q}) \operatorname{Re}(Y_{ab}(\mathbf{q},\omega))$$

$$= -F_{ab}\Delta_{ab} + \Lambda_{ab} \frac{|v_{ab}|^2}{|D|^2} \sum_{c \neq a,b} \Lambda_{cc} v_{cc} \operatorname{Im}\left(\Pi_{aa}\Pi^*_{bb}\Pi''_{cc} f_c - \Pi^*_{cc}\Pi_{aa}\Pi''_{bb} f_b - \Pi_{cc}\Pi^*_{bb}\Pi''_{aa} f_a\right)$$

(99)

$$= -F_{ab}\Delta_{ab} + \Lambda_{ab} \frac{|v_{ab}|^2}{|D|^2} \sum_{c \neq a,b} \Lambda_{cc} v_{cc} \left(\Pi''_{aa}\Pi''_{bb}\Pi'_{cc}\Delta_{ab} + \Pi''_{bb}\Pi''_{cc}\Pi'_{aa}\Delta_{bc} + \Pi''_{cc}\Pi''_{aa}\Pi'_{bb}\Delta_{ca}\right)$$

$$= -F_{ab}\Delta_{ab} + \frac{1}{\Lambda_{ab}} \sum_c \Delta_{abc}$$

where

$$\Delta_{abc} = \frac{1}{|D|^2} \Lambda_{aa} v_{aa} \Lambda_{bb} v_{bb} \Lambda_{cc} v_{cc} \left(\Pi''_{aa}\Pi''_{bb}\Pi'_{cc}\Delta_{ab} + \Pi''_{bb}\Pi''_{cc}\Pi'_{aa}\Delta_{bc} + \Pi''_{cc}\Pi''_{aa}\Pi'_{bb}\Delta_{ca}\right) \quad (100)$$

An essential property of $\Delta_{abc}$ is that it is antisymmetric with respect to interchange of any two indicies, eg $\Delta_{abc} = -\Delta_{bac} = \Delta_{bca}$ etc, and consequently vanishes if any two are equal. This enables the restriction $c \neq a,b$ on the summation index on the right-hand side of (99) to be lifted, and the sum effectively extended over all the component species present. Note also that $\phi_{ab}$ given by (99) is antisymmetric, ie $\phi_{ab} = -\phi_{ba}$ as required for overall energy conservation.

In general, the right-hand side of (99) contains non-vanishing cross-coupling terms involving the spectator species. These terms are a consequence of the persistence of the higher order terms in (46) by which the 'direct' energy transfer between the specified species is affected by the deviation from equilibrium with other species present. From a physical point of view, it is not meaningful to ask precisely how energy flows from one species to another, when all the species are closely coupled within the same volume of space. One can only meaningfully ask how the energy of any one species changes in time. So, in the case of species a, for example,

$$\dot{\mathcal{E}}_a = \sum_b \dot{\mathcal{E}}_{ba} = -\sum_b \dot{\mathcal{E}}_{ab} \quad (101)$$

which can be expressed by

$$\dot{\mathcal{E}}_a = \frac{1}{(2\pi)^4} \iint_{-\infty}^{+\infty} \phi_a(\mathbf{q},\omega) \, d\omega \, d^3\mathbf{q} \quad (102)$$

where

$$\phi_a = \sum_b \phi_{ba} = -\sum_b \phi_{ab} \quad (103)$$

Combining (99) and (103) yields,

$$\phi_a = \sum_b (F_{ab}\Delta_{ab}) - \frac{1}{\Lambda_{ab}}\sum_{b,c}\Delta_{abc} \qquad (104)$$

Next we use the property that the $\Delta$s with two indices represent simple differences (so that cyclic sums, like $\Delta_{ab} + \Delta_{bc} + \Delta_{ca}$ always vanish) to eliminate $\Delta_{bc}$, according to $\Delta_{bc} = \Delta_{ba} + \Delta_{ac}$, to yield

$$\Delta_{abc} = \frac{1}{|D|^2}\Lambda_{aa}v_{aa}\Lambda_{bb}v_{bb}\Lambda_{cc}v_{cc}\left((\Pi''_{aa}\Pi'_{cc} - \Pi''_{cc}\Pi'_{aa})\Pi''_{bb}\Delta_{ab} + (\Pi''_{bb}\Pi'_{aa} - \Pi''_{aa}\Pi'_{bb})\Pi''_{cc}\Delta_{ac}\right) \qquad (105)$$

and hence

$$\phi_a = \sum_b (F_{ab}\Delta_{ab}) - \frac{\Lambda_{aa}v_{aa}}{|D|^2}\left(\begin{array}{l}\displaystyle\sum_{b,c}\frac{\Lambda_{bb}v_{bb}\Lambda_{cc}v_{cc}}{\Lambda_{ab}}(\Pi''_{aa}\Pi'_{cc} - \Pi''_{cc}\Pi'_{aa})\Pi''_{bb}\Delta_{ab} \\ \displaystyle -\sum_{b,c}\frac{\Lambda_{bb}v_{bb}\Lambda_{cc}v_{cc}}{\Lambda_{ab}}(\Pi''_{aa}\Pi'_{bb} - \Pi''_{bb}\Pi'_{aa})\Pi''_{cc}\Delta_{ac}\end{array}\right)$$

$$= \sum_b (F_{ab}\Delta_{ab}) - \frac{\Lambda_{aa}v_{aa}}{|D|^2}\sum_{b,c}\Lambda_{bb}v_{bb}\Lambda_{cc}v_{cc}(\Pi''_{aa}\Pi'_{cc} - \Pi''_{cc}\Pi'_{aa})\Pi''_{bb}\Delta_{ab}\left(\frac{1}{\Lambda_{ab}} - \frac{1}{\Lambda_{ac}}\right) \qquad (106)$$

$$= \sum_b \tilde{F}_{ab}\Delta_{ab}$$

where

$$\tilde{F}_{ab} = F_{ab} - \frac{\Lambda_{aa}v_{aa}\Lambda_{bb}v_{bb}}{|D|^2}\Pi''_{bb}\sum_c \Lambda_{cc}v_{cc}(\Pi''_{aa}\Pi'_{cc} - \Pi''_{cc}\Pi'_{aa})\left(\frac{1}{\Lambda_{ab}} - \frac{1}{\Lambda_{ac}}\right)$$

$$= F_{ab}\left(1 - \sum_c \Lambda_{cc}v_{cc}\left(\Pi'_{cc} - \Pi'_{aa}\frac{\Pi''_{cc}}{\Pi''_{aa}}\right)\left(1 - \frac{\Lambda_{ab}}{\Lambda_{ac}}\right)\right) \qquad (107)$$

$$= F_{ab}\left(1 - \sum_c \chi'_{cc}\left(1 - \frac{\Pi'_{aa}\Pi''_{cc}}{\Pi''_{aa}\Pi'_{cc}}\right)\left(1 - \frac{\Lambda_{ab}}{\Lambda_{ac}}\right)\right)$$

The final result for a multi-component plasma, is therefore

$$\dot{\mathcal{E}}_a = \frac{1}{(2\pi)^4}\iint_{-\infty}^{+\infty}\sum_b \tilde{F}_{ab}(\mathbf{q},\omega)\Delta_{ab}(\omega)\,d\omega\,d^3\mathbf{q}$$

$$= \frac{1}{(2\pi)^4}\sum_b \iint_{-\infty}^{+\infty}\tilde{F}_{ab}(\mathbf{q},\omega)\Delta_{ab}(\omega)\,d\omega\,d^3\mathbf{q} \qquad (108)$$

# 4 STATIC LOCAL FIELD APPROXIMATION FOR COULOMB SYSTEMS

## 4.1 Static correlations and local field corrections

The local field corrections $\Lambda_{ab}(\mathbf{q})$ represent the local corrections to the static potential that take account of short-range correlations between the particles. The general result expressed by (87) for the electron-ion exchange rate depends explicitly on the electron-ion correlation correction, and only weakly, through the quantum-plasma correction $\Delta(\mathbf{q})$, on the other elements of $\Lambda$.

Methods for calculating LFCs can be found in the literature. These typically rely on classical approximations, such as Hypernetted Chain (HNC) or Percus-Yevik (PY). In the following, we give a first-principles approach in which classical approximations are relied upon only for the ion-ion correlations, for which they are generally considered appropriate.

The static correlations in a many body system are expressed by the pair correlation functions $g_{ab}(\mathbf{r})$ which are given in terms of the static response to the effective particle interaction, in the first instance, as follows

$$n_a \int e^{-i\mathbf{q}\cdot\mathbf{r}} g_{ab}(\mathbf{r}) d^3\mathbf{r} = \sum_c K_{ac}(\mathbf{q},0) v_{cb}^{\text{eff}}(\mathbf{q}) \tag{109}$$

which holds for Boltzmann particles in both classical and quantum regimes. Equation (109) gives the density response to the electrostatic potential of a source particle, $b$. In the absence of any potential, ie if the source particle does not interact via a potential with any other particles in the system, then there is no response and the particle is effectively invisible to the system. If however the source particles $b$ give rise to exchange correlations, due to the presence of other indistinguishable fermion particles in the system, then this becomes modified, by exchange correlations, to

$$n_a \int e^{-i\mathbf{q}\cdot\mathbf{r}} \left(\tilde{g}_{ab}(\mathbf{r}) - \delta_{ab} g_{bb}^0(\mathbf{r})\right) d^3\mathbf{r} = \left(1 + n_b \int e^{-i\mathbf{q}\cdot\mathbf{r}} g_{bb}^0(\mathbf{r}) d^3\mathbf{r}\right)\left(\sum_c K_{ac}(\mathbf{q},0) v_{cb}^{\text{eff}}(\mathbf{q})\right) \tag{110}$$

(see APPENDIX B) which is equivalent to

$$\tilde{g}_{ab}(\mathbf{q}) \equiv \sqrt{n_a n_b} \int e^{-i\mathbf{q}\cdot\mathbf{r}} \tilde{g}_{ab}(\mathbf{r}) d^3\mathbf{r} = \delta_{ab} g_{aa}^0(\mathbf{q}) + \sqrt{\frac{n_b}{n_a}}\left(1 + g_{bb}^0(\mathbf{q})\right) \sum_c K_{ac}(\mathbf{q},0) v_{cb}^{\text{eff}}(\mathbf{q}) \tag{111}$$

in which $g_{aa}^0$ is the (Lindhard) exchange pair correlation function for non-interacting indistinguishable fermion particles and $\tilde{g}_{ab}(\mathbf{q})$ is the intermediate (non-symmetric) pair correlation matrix for interacting particles (in terms of which, the true PCF, $g_{ab}(\mathbf{q})$ is later shown later to be given by (218) - (219)). These equations include the quantum corrections accounting for quantum exchange correlations. While $\tilde{g}_{aa}(\mathbf{q}) = g_{aa}(\mathbf{q})$, in general, for $a \neq b$, $\tilde{g}_{ab}(\mathbf{q}) \neq \tilde{g}_{ba}(\mathbf{q})$. However, if the particles are in thermal equilibrium, ie $T_a = T_b$, then $\tilde{g}_{ab}(\mathbf{0}) = \tilde{g}_{ba}(\mathbf{0}) = g_{ab}(\mathbf{0})$, ie the long wavelength limit of $\tilde{g}_{ab}$ is symmetric and equal to the long-wavelength limit of $g_{ab}$. Otherwise, for $a \neq b$, $\tilde{g}_{ab}$ may be regarded as an approximation to $g_{ab}$ in the following cases: (i) for classical systems and (ii) for systems for which $m_a \ll m_b$, provided that the temperatures of the particles are such that the mean square velocity of the heavier particles is much smaller than that of the lighter particles.

It follows from equation (14) that the static response $\mathbf{K}(\mathbf{q},0)$ is Hermitian, and hence, from (49),

$$\mathrm{K}_{ab}(\mathbf{q},0) = \Pi^0_{aa}(\mathbf{q},0)\delta_{ab} + \Pi^0_{aa}(\mathbf{q},0)\sum_c v^{\mathrm{eff}}_{ac}(\mathbf{q})\mathrm{K}_{cb}(\mathbf{q},0)$$

$$= \Pi^0_{bb}(\mathbf{q},0)\delta_{ab} + \Pi^0_{bb}(\mathbf{q},0)\sum_c \mathrm{K}_{ac}(\mathbf{q},0)v^{\mathrm{eff}}_{cb}(\mathbf{q})$$

(112)

Moreover, it follows from (5) that $g_{ab}(\mathbf{q})$ and hence $\tilde{g}_{ab}(\mathbf{q})$ and $\mathrm{K}_{ab}(\mathbf{q},0)$ are real. The matrix $\mathbf{K}(\mathbf{q},0)$ is therefore real and symmetric.

Upon combining (111) and (112), an explicit closed formula for the intermediate PCF is

$$\tilde{g}_{ab}(\mathbf{q}) = \left(1 + g^0_{bb}(\mathbf{q})\right)\sqrt{\frac{n_b}{n_a}}\frac{\mathrm{K}_{ab}(\mathbf{q},0)}{\Pi^0_{bb}(\mathbf{q},0)} - \delta_{ab} \quad (113)$$

Now let

$$\mathfrak{s}_{ab}(\mathbf{q}) = \delta_{ab} + \tilde{g}_{ab}(\mathbf{q}) \quad (114)$$

which is equivalent to

$$\mathfrak{s} = \mathbf{1} + \tilde{\mathbf{g}} \quad (115)$$

where $\mathfrak{s} = \|\mathfrak{s}_{ab}(\mathbf{q})\|$ and

$$\mathfrak{s}_{ab}(\mathbf{q}) = \sqrt{\frac{n_b}{n_a}}\frac{\mathrm{K}_{ab}(\mathbf{q},0)}{\Pi^0_{bb}(\mathbf{q},0)}\left(1 + g^0_{bb}(\mathbf{q})\right) \equiv \sqrt{\frac{n_b}{n_a}}\sigma_{ab}(\mathbf{q})\left(1 + g^0_{bb}(\mathbf{q})\right) \quad (116)$$

where, by virtue of equation (112), the matrix $\boldsymbol{\sigma} = \|\sigma_{ab}(\mathbf{q})\| \equiv \left\|\frac{\mathrm{K}_{ab}(\mathbf{q},0)}{\Pi^0_{bb}(\mathbf{q},0)}\right\|$ satisfies

$$\boldsymbol{\sigma} = \mathbf{1} + \boldsymbol{\sigma}\boldsymbol{\chi}_0 \quad (117)$$

in which $\boldsymbol{\chi}_0 = \|\chi_{ab}(\mathbf{q},0)\|$ is the static polarizability matrix as per (51). It follows directly that

$$\boldsymbol{\sigma} = (1-\boldsymbol{\chi}_0)^{-1} \tag{118}$$

and hence the components of $\mathfrak{s}$ are given by, making use of (23),

$$\mathfrak{s}_{ab}(\mathbf{q}) = (1+g_{bb}^0(\mathbf{q}))\sqrt{\frac{n_b}{n_a}}(1-\boldsymbol{\chi}_0)^{-1}\bigg|_{ab}$$

$$= \left(\delta_{ab} + \sqrt{\frac{n_b}{n_a}}\frac{\chi_{ab}(\mathbf{q},0)}{D(\mathbf{q},0)}\right)(1+g_{bb}^0(\mathbf{q})) \tag{119}$$

the inverse of which yields

$$\mathfrak{s}^{-1}\bigg|_{ab} = (1+g_{aa}^0(\mathbf{q}))^{-1}\left(\delta_{ab} - \sqrt{\frac{n_b}{n_a}}\chi_{ab}(\mathbf{q},0)\right) \tag{120}$$

Introducing the quantum direct correlation matrix,

$$\mathfrak{c} \equiv \|c_{ab}(\mathbf{q})\| = \left\|\frac{1}{1+g_{aa}^0(\mathbf{q})}\left(g_{aa}^0(\mathbf{q})\delta_{ab} + \sqrt{\frac{n_b}{n_a}}\chi_{ab}(\mathbf{q},0)\right)\right\|$$

$$= \left\|\frac{1}{1+g_{aa}^0(\mathbf{q})}\left(g_{aa}^0(\mathbf{q})\delta_{ab} + \sqrt{\frac{n_b}{n_a}}\Pi_{aa}^0(\mathbf{q},0)v_{ab}^{\text{eff}}(\mathbf{q})\right)\right\| \tag{121}$$

in terms of which, the LFCs are given by

$$\Lambda_{ab}(\mathbf{q}) = \sqrt{\frac{n_a}{n_b}}\frac{(1+g_{aa}^0(\mathbf{q}))c_{ab}(\mathbf{q}) - g_{aa}^0(\mathbf{q})\delta_{ab}}{\Pi_{aa}^0(\mathbf{q},0)v_{ab}(\mathbf{q})} = \frac{\chi_{ab}(\mathbf{q},0)}{\Pi_{aa}^0(\mathbf{q},0)Q_aQ_bv(\mathbf{q})} \tag{122}$$

equations (119) and (120) respectively become

$$\mathfrak{s}_{ab}(\mathbf{q}) = \delta_{ab}(1+g_{aa}^0(\mathbf{q}))\left(1-\frac{g_{aa}^0(\mathbf{q})}{D(\mathbf{q},0)}\right) + (1+g_{aa}^0(\mathbf{q}))\frac{c_{ab}(\mathbf{q})}{D(\mathbf{q},0)}(1+g_{bb}^0(\mathbf{q})) \tag{123}$$

$$\mathfrak{s}^{-1} = 1-\mathfrak{c} \tag{124}$$

and hence

$$\mathfrak{s} = 1+\mathfrak{c}\mathfrak{s} \tag{125}$$

Expressing $\mathfrak{s}$ in (125) in terms of the intermediate pair correlation matrix $\tilde{\mathfrak{g}}$ in accordance with (115) yields the general quantum Ornstein-Zernicke (QOZ) relation

$$\tilde{\mathfrak{g}} = \mathfrak{c} + \mathfrak{c}\tilde{\mathfrak{g}} \tag{126}$$

the solution of which is

$$\tilde{g}_{ab}(\mathbf{q}) \equiv \mathfrak{s}_{ab}(\mathbf{q}) - \delta_{ab} = \delta_{ab} g_{aa}^0(\mathbf{q})\left(1 - \frac{1 + g_{aa}^0(\mathbf{q})}{D(\mathbf{q},0)}\right) + \left(1 + g_{aa}^0(\mathbf{q})\right)\frac{c_{ab}(\mathbf{q})}{D(\mathbf{q},0)}\left(1 + g_{bb}^0(\mathbf{q})\right)$$

(127)

$$= \delta_{ab} g_{aa}^0(\mathbf{q}) + \sqrt{\frac{n_b}{n_a}}\frac{\chi_{ab}(\mathbf{q},0)}{D(\mathbf{q},0)}\left(1 + g_{bb}^0(\mathbf{q})\right)$$

which follows from (123) and (115), where $D$ can be expressed variously in terms of either $\mathfrak{c}$ or $\mathfrak{g}$ as follows,

$$D(\mathbf{q},0) = 1 - \sum_a \left(c_{aa}\left(1 + g_{aa}^0\right) - g_{aa}^0\right)$$

(128)

$$\frac{1}{D(\mathbf{q},0)} = 1 + \sum_a \frac{g_{aa} - g_{aa}^0}{1 + g_{aa}^0} = 1 + \text{trace}\left(\mathfrak{g}'(\mathbf{q})\right)$$

where, $\mathfrak{g}'$ is the Boltzmann PCF defined by

$$\mathfrak{g}' = \left\|\frac{\tilde{g}_{ab} - \delta_{ab}g_{aa}^0}{1 + g_{bb}^0}\right\| = \left\|\sqrt{\frac{n_b}{n_a}}\frac{\chi_{ab}(\mathbf{q},0)}{D(\mathbf{q},0)}\right\|$$

(129)

and which satisfies (109). Since $1/D \geq 0$, the second of these equations implies, for Coulomb systems, the interesting and non-trivial inequality

$$\text{trace}\left(\mathfrak{g}'(\mathbf{q})\right) \geq -1$$

(130)

The matrices $\tilde{\mathfrak{g}}$, $\mathfrak{g}'$, $\mathfrak{s}$, $\mathfrak{c}$ and $\chi_0$ are not symmetric. Referring to (113) - (116) and (121), one finds

$$\frac{\tilde{g}_{ab}(\mathbf{q})}{\tilde{g}_{ba}(\mathbf{q})} = \frac{\mathfrak{s}_{ab}(\mathbf{q})}{\mathfrak{s}_{ba}(\mathbf{q})} = \frac{c_{ab}(\mathbf{q})}{c_{ba}(\mathbf{q})} = \frac{n_b \Pi_{aa}^0(\mathbf{q},0)\left(1 + g_{bb}^0(\mathbf{q})\right)}{n_a \Pi_{bb}^0(\mathbf{q},0)\left(1 + g_{aa}^0(\mathbf{q})\right)}$$

(131)

$$\frac{g'_{ab}(\mathbf{q})}{g'_{ba}(\mathbf{q})} = \frac{n_b \chi_{ab}(\mathbf{q},0)}{n_a \chi_{ba}(\mathbf{q},0)} = \frac{n_b \Pi_{aa}^0(\mathbf{q},0)}{n_a \Pi_{bb}^0(\mathbf{q},0)}$$

Moreover the elements of each of the matrices $\tilde{\mathfrak{g}} - \mathfrak{g}_0$, $\mathfrak{g}'$ and $\chi_0$ all satisfy the general relation,

$$X_{ab}X_{ba} - X_{aa}X_{bb} = 0, \quad \forall a,b$$

(132)

which is a consequence of the Coulomb nature of the underlying interactions. In terms of the functions $R_a(\mathbf{q})$ and $\tilde{R}_a(\mathbf{q})$, which, following ref. [28], may be defined by

$$\tilde{R}_a(\mathbf{q}) = \frac{2T_a}{\pi n_a v_{aa}(\mathbf{q})} \int_0^\infty \mathrm{Im}\,\varepsilon_a^0(\mathbf{q},\omega)\frac{\mathrm{d}\omega}{\omega}$$

$$R_a(\mathbf{q}) = \frac{2T_a}{\pi n_a v_{aa}(\mathbf{q})} \int_0^\infty \frac{\omega}{2T_a}\coth\left(\frac{\omega}{2T_a}\right)\mathrm{Im}\,\varepsilon_a^0(\mathbf{q},\omega)\frac{\mathrm{d}\omega}{\omega}$$

(133)

where $\varepsilon_a^0(\mathbf{q},\omega)$ is the RPA dielectric function for particles of species $a$ interacting via the Coulomb potential $v_{aa}(\mathbf{q})$, the Lindhard PCF and susceptibility are given by

$$1 + g_{aa}^0(\mathbf{q}) = R_a(\mathbf{q}) \tag{134}$$

$$\Pi_{aa}^0(\mathbf{q},0) = -\frac{n_a}{T_a}\tilde{R}_a(\mathbf{q}) \tag{135}$$

This yields

$$\frac{\tilde{g}_{ab}(\mathbf{q})}{\tilde{g}_{ba}(\mathbf{q})} = \frac{s_{ab}(\mathbf{q})}{s_{ba}(\mathbf{q})} = \frac{c_{ab}(\mathbf{q})}{c_{ba}(\mathbf{q})} = \frac{T_b\tilde{R}_a(\mathbf{q})R_b(\mathbf{q})}{T_a\tilde{R}_b(\mathbf{q})R_a(\mathbf{q})}$$

$$\frac{g'_{ab}(\mathbf{q})}{g'_{ba}(\mathbf{q})} = \frac{n_b\chi_{ab}(\mathbf{q},0)}{n_a\chi_{ba}(\mathbf{q},0)} = \frac{T_b\tilde{R}_a(\mathbf{q})}{T_a\tilde{R}_b(\mathbf{q})}$$

(136)

The properties of the functions $R_a(\mathbf{q})$ and $\tilde{R}_a(\mathbf{q})$ are discussed in detail elsewhere [28], and can be summarised as follows: both functions tend to the same limit, denoted by $R_{a0}$, in the limit of $\mathbf{q}=\mathbf{0}$; and both tend to unity for Boltzmann particles in the classical limit. In the limit of large $q=|\mathbf{q}|$, $\tilde{R}_a(\mathbf{q}) \sim q^{-2}$ and $\lim_{q\to\infty} R_a(\mathbf{q})=1$. Moreover, since $\mathrm{Im}\,\varepsilon_a^0(\mathbf{q},\omega) \propto v_{aa}(\mathbf{q})$, the functions are independent of the strength of the Coulomb interaction between the particles and are therefore exactly given by the RPA in the limit when the particles are non-interacting, ie $Q_a^2 \to 0$. The function $R_a(\mathbf{q})$, for example, is precisely the RPA static structure factor for non-interacting particles, and describes pure exchange correlations. In the classical limit, for example, we then have [5],

$$\frac{\tilde{g}_{ab}(\mathbf{q})}{\tilde{g}_{ba}(\mathbf{q})} \sim \frac{T_b}{T_a} \tag{137}$$

while, for small $\mathbf{q}$,

$$\frac{\tilde{g}_{ab}(\mathbf{q})}{\tilde{g}_{ba}(\mathbf{q})} = \frac{T_b}{T_a}\left(1 + \frac{q^2}{12}\left(\frac{1}{m_b T_b R_{b0}} - \frac{1}{m_a T_a R_{a0}}\right) + \mathcal{O}(q^4)\right) \tag{138}$$

A procedure for symmetrising these matrices to determine the proper PCF, $g_{ab}(\mathbf{q})$, in the presence of quantum exchange corrections, is given in APPENDIX B.

From equation (115), the diagonal elements of the matrix $\mathbf{s}$ can be seen to be just the static structure factors for each separate component species. The general element

$$S_{ab}(\mathbf{q}) = \int_{-\infty}^{+\infty} S_{ab}(\mathbf{q},\omega)\,d\omega = \int_{-\infty}^{+\infty} \widehat{S}_{ab}(\mathbf{q},\omega)\,d\omega \qquad (139)$$

of the static structure factor is real and symmetric and is related to $s_{ab}(\mathbf{q})$, by the proposed symmetrization procedure described in APPENDIX B, according to

$$S_{ab}(\mathbf{q}) = \left(m_a \tilde{R}_a(\mathbf{q}) + m_b \tilde{R}_b(\mathbf{q})\right)\left(\frac{m_b \tilde{R}_b(\mathbf{q})}{s_{ab}(\mathbf{q})} + \frac{m_a \tilde{R}_a(\mathbf{q})}{s_{ba}(\mathbf{q})}\right)^{-1} \qquad (140)$$

Combining (122) and (127) expresses the LFCs directly in terms of the pair correlation functions as follows

$$\Lambda_{ab}(\mathbf{q}) = \sqrt{\frac{n_a}{n_b}} \frac{g'_{ab}(\mathbf{q})D(\mathbf{q},0)}{\Pi^0_{aa}(\mathbf{q},0)v_{ab}(\mathbf{q})} \qquad (141)$$

in which $D(\mathbf{q},0)$, the static dielectric function, is provided by the second of equations (128), and $g'_{ab}(\mathbf{q})$ is defined by (129). Using (129) and (131), it can readily be shown that $\Lambda$ given by (141) is symmetric, ie $\Lambda_{ab} = \Lambda_{ba}$, and, by referring to (75), satisfies (53).

The pair correlation matrix can be determined by using a combination of techniques including DFT, classical MD and closure relations such as the classical HNC or PY approximations. Classical methods are generally applicable to heavy particle systems (eg, to determine the ion-ion correlations) while quantal methods, such as DFT, should generally be used to describe the electrons. Once this is done, the LFCs are provided by equation (141).

By way of illustration, a practical procedure for treating the two-component plasma is proposed as follows.

A classical treatment is applied to the ions (no exchange). Using an initial guess for the function $\Sigma(\mathbf{r})$ (eg $\Sigma(\mathbf{r})=0$) in the first instance, the ion-ion PCF $g_{ii}(\mathbf{r})$ is obtained from the general closure relation [20].

$$g_{ii}(\mathbf{r}) = \exp\left(-\frac{v_{ii}(\mathbf{r})}{T_i} + \Sigma(\mathbf{r})\right) - 1 \qquad (142)$$

where, in the HNC approximation,

$$\Sigma(\mathbf{r}) = \Sigma_0(\mathbf{r}) \qquad (143)$$

or, in the PY approximation,

$$e^{\Sigma(\mathbf{r})} = 1 + \Sigma_0(\mathbf{r}) \qquad (144)$$

where

$$\Sigma_0(\mathbf{r}) = \frac{1}{(2\pi)^3 n_i} \int (g_{ii}(\mathbf{q}) - c_{ii}(\mathbf{q})) e^{i\mathbf{q}\cdot\mathbf{r}} d^3\mathbf{q}$$

$$= \frac{1}{(2\pi)^3 n_i} \int (1 - D(\mathbf{q},0)) g_{ii}(\mathbf{q}) e^{i\mathbf{q}\cdot\mathbf{r}} d^3\mathbf{q}$$

(145)

in which the Fourier transforms are defined as per (111). Next determine the electron-ion correlation function $g_{ei}^*$ by means of a single-centre atomic cell DFT or average-atom calculation. The complete electron-ion correlation function is then given by a convolution of $g_{ei}^*$ and $g_{ii}$ as follows:

$$g_{ei}(\mathbf{q}) = g_{ei}^*(\mathbf{q}) + g_{ii}(\mathbf{q}) g_{ei}^*(\mathbf{q})$$

$$= (1 + g_{ii}(\mathbf{q})) g_{ei}^*(\mathbf{q})$$

(146)

Details of the general formulation of this result are given in section 4.2. Since the correlation matrix is symmetric, the ion-electron correlation function is $g_{ie}(\mathbf{q}) = g_{ei}(\mathbf{q})$. Moreover, since $m_e \ll m_i$ and provided that the temperatures are such that $\langle v_e^2 \rangle \gg \langle v_i^2 \rangle$, which means that the ion temperature should not be too much higher than the electron temperature, the intermediate electron-ion correlation function is determined by $\tilde{g}_{ei}(\mathbf{q}) \simeq g_{ei}(\mathbf{q})$. The corresponding intermediate ion-electron correlation function is then given, according to (131), by:

$$\tilde{g}_{ie}(\mathbf{q}) = \frac{n_e}{n_i} \frac{\Pi_{ii}^0(\mathbf{q},0)}{\Pi_{ee}^0(\mathbf{q},0)} \frac{1 + g_{ee}^0(\mathbf{q})}{1 + g_{ii}^0(\mathbf{q})} \tilde{g}_{ei}(\mathbf{q}) \simeq -\frac{n_e (1 + g_{ee}^0(\mathbf{q}))}{T_i \Pi_{ee}^0(\mathbf{q},0)} g_{ei}(\mathbf{q})$$

$$= \frac{T_e}{T_i} \frac{R_e(\mathbf{q})}{\tilde{R}_e(\mathbf{q})} g_{ei}(\mathbf{q})$$

(147)

The electron-electron correlation function is then provided by applying (132) to $\tilde{\mathfrak{g}} - \mathfrak{g}_0$, to yield

$$g_{ee}(\mathbf{q}) - g_{ee}^0(\mathbf{q}) = \frac{\tilde{g}_{ei}(\mathbf{q})\tilde{g}_{ie}(\mathbf{q})}{g_{ii}(\mathbf{q})}$$

$$\simeq -\frac{n_e\left(1 + g_{ee}^0(\mathbf{q})\right)}{T_i \Pi_{ee}^0(\mathbf{q},0)} \frac{g_{ei}(\mathbf{q})^2}{g_{ii}(\mathbf{q})} \qquad (148)$$

$$= \frac{T_e}{T_i} \frac{R_e(\mathbf{q})}{\tilde{R}_e(\mathbf{q})} \frac{g_{ei}(\mathbf{q})^2}{g_{ii}(\mathbf{q})}$$

Then, from (128),

$$1/D(\mathbf{q},0) = 1 + g_{ii}(\mathbf{q}) + \frac{g_{ee}(\mathbf{q}) - g_{ee}^0(\mathbf{q})}{1 + g_{ee}^0(\mathbf{q})}$$

$$\simeq 1 + g_{ii}(\mathbf{q}) - \frac{n_e}{T_i \Pi_{ee}^0(\mathbf{q},0)} \frac{g_{ei}(\mathbf{q})^2}{g_{ii}(\mathbf{q})} \qquad (149)$$

$$= \left(1 + g_{ii}(\mathbf{q})\right)\left(1 + \frac{T_e}{T_i}\left(\frac{1 + g_{ii}(\mathbf{q})}{g_{ii}(\mathbf{q})}\right) \frac{g_{ei}^*(\mathbf{q})^2}{\tilde{R}_e(\mathbf{q})}\right)$$

Equations (142) - (149) have to be solved iteratively.

Finally, the LFCs can be calculated according to (141).

Setting $\Lambda_{ab} = 1 \ \forall a,b$ and using the classical approximation $\Pi_{aa}^0 \simeq -n_a/T_a$ for all particles, leads to the classical Debye-Hückel approximation, $D(\mathbf{q},0) = 1 + \frac{k^2}{q^2}$, with $k^2 = \sum_a k_a^2$, $k_a^2 = \frac{Q_a^2 e^2 n_a}{\varepsilon_0 T_a}$ as can be readily verified, while the pair correlation function becomes $g_{ab}(\mathbf{q}) = -\frac{Q_b}{Q_a}\sqrt{\frac{n_b}{n_a}} \frac{k_a^2}{k^2 + q^2}$, which is the Debye-Hückel PCF in agreement with formula given in [28], for example. With LFCs included, the classical static effective dielectric function becomes $D(\mathbf{q},0) = 1 + \frac{1}{q^2}\sum_a k_a^2 \Lambda_{aa}(\mathbf{q})$. It should be borne in mind that these functions describe the correlations between the constituents of a neutral homogeneous plasma, which may not be quite the same as the charge distribution around a test charge introduced into an otherwise neutral plasma. The latter creates a system which is neither electrically neutral and nor strictly homogeneous. The electric field implied by the inherent non-neutrality cannot be represented on the 3-torus manifold of a homogeneous system in 3 dimensions with cyclic boundary conditions.

## 4.2 Bound electrons and relationship to single-centre atomic calculations

The above is applicable, in the first instance, to fully-ionised neutral systems comprising bare atomic nuclei and delocalized (continuum) electrons. If the plasma contains ions in which some electrons are bound to nuclei, then it might be tempting to treat each of these partially-ionised ions as a separate particle species, ie to adopt the chemical picture [16]. In general, this would be wrong for two reasons: (i). the existence of transient bound states, which are internal states in the plasma that are dynamically populated, by recombination, and depopulated, by ionization, would mean that local continuity, as expressed by equation (42), is violated. (ii) The screening produced by bound electrons will give rise to deviations from the assumed Coulomb interactions, which require that the interaction between any two particles is Coulombic and proportional to the product of the (fixed) charges. For these reasons, the model as described is essentially an all-electron one, requiring that one should adopt the physical picture in which the constituent particles are conserved and interact solely via Coulomb forces. However approximations may be made in which permanently bound core electrons, provided they are confined to volumes that are, in total, very small compared with the total volume of the plasma, are excluded from the calculation and included with the nuclear charge. The resulting pseudopotentials must however remain Coulomb in nature.

A further point to consider is the effect of bound-state and resonance dynamics on the electron-ion interaction. Such effects are beyond the static local field approximation proposed here.

In weakly-coupled classical plasmas, the upper bound levels are very weakly populated with the majority of the bound electrons localized within a volume that is small compared with the ion-sphere volume. In such situations, it may be justifiable to use an average-ion description, ie, to treat all of the ions of a particular nuclear species as comprising a single ion species to which an average charge is assigned.

In general however, one should employ an all electron approach in which all the electrons are treated on an equal footing.

All-electron treatments of complex atoms typically involve modelling a single average or typical atom in an idealised plasma environment treated as a non-fluctuating continuous distribution of electrons immersed in a neutralising positively charged background. This is sometimes referred to as the atom-in-jellium model [30], [31] an essential feature of which is that the charges making up the environment are correlated only with the central charge. The jellium is therefore not a representation of the surrounding plasma *per se*, but rather a representation of its effect on the subject atom. An ensemble of atoms can be treated by considering each atom to be independently immersed in the jellium, which is characterised by a common temperature and chemical potential, and in thermodynamic equilibrium with it. The atoms are consequently in thermodynamic equilibrium with each other. If the jellium is then subtracted out, and the atoms brought together, one is left with a realistic model of the real plasma. In this sense, the jellium is like scaffolding, which allows states of individual atoms within a plasma to be modelled without having to model, in comparable detail, the entire plasma around them. In the atom-in-jellium model, the electron distribution associated with an atom comprises three components: bound electrons, free electrons closely coupled to the core ion in sufficient number to make the system neutral and electrons associated with the surrounding plasma, which is considered to extend to infinity. The electron distribution surrounding the central nucleus would then be expressed by

$$\rho_{ei}^*(\mathbf{r}) = n_i \left( Z_b g_{ei}^b(\mathbf{r}) + Z_f g_{ei}^f(\mathbf{r}) + Z^* \left(1 + g_0^*(\mathbf{r})\right)\right)$$

$$\equiv Z n_i \left( \frac{Z^*}{Z}\left(1+ g_0^*(\mathbf{r})\right) + g_{ei}^*(\mathbf{r}) \right)$$

(150)

where $n_b = Z_b n_i$ is the bound electron density, $n_f = Z_f n_i$ is the free electron density and $Z_f + Z_b = Z$, and where $n_e^* = Z^* n_i$ is the effective average electron density, which is the far-field free electron density in the surrounding jellium, in terms of which the degeneracy parameter of the plasma would be given by (88). The pair correlation functions $g_{ei}^f(\mathbf{r})$ and $g_{ei}^b(\mathbf{r})$ for the free and bound electrons respectively are defined to vanish for $r \to \infty$ and possess volume integrals as follows

$$n_i \int g_{ei}^f(\mathbf{r}) d^3\mathbf{r} = 1$$

$$n_i \int g_{ei}^b(\mathbf{r}) d^3\mathbf{r} = 1$$

(151)

The electron distribution in the jellium is represented by $\rho_0^*(\mathbf{r}) = Z^* n_i \left(1 + g_0^*(\mathbf{r})\right)$, where the 'jellium correlation function' $g_0^*(\mathbf{r})$ is a provisional function with the essential properties that $g_0^*(\mathbf{0}) = -1$ and $g_0^*(\infty) = 0$. The effective charge $Z^*$ is determined by continuity of the total charge distribution under equilibrium conditions. Polarization of the electron density by the central ion generally implies that $Z^* < Z_f$. In this model, overall electrical neutrality implies a background positive charge distribution $\sim \rho_0^*(\mathbf{r})$ which thus associates $Z^*$ free electrons with each ion in the surrounding plasma. In the ubiquitous ion-sphere model of a strongly-coupled plasma, $g_0^*(\mathbf{r})$ is the step function,

$$g_0^*(\mathbf{r}) = -1 : r < R_0$$

$$= 0 : r > R_0$$

(152)

where $R_0 = (3/4\pi n_i)^{1/3}$ is the ion-sphere radius. A more general form of this function, which accommodates weakly coupled plasmas, is suggested elsewhere [32], [33], [34]. It should be noted that the final correlation functions do not depend directly on $g_0^*(\mathbf{r})$, but only indirectly through the effects on the atomic model.

The single-centre electron-ion PCF defined by (150) is

$$g_{ei}^*(\mathbf{r}) = \frac{1}{Z}\left(Z_f g_{ei}^f(\mathbf{r}) + Z_b g_{ei}^b(\mathbf{r})\right) \tag{153}$$

which, according to (151), possesses the volume integral

$$n_i \int g_{ei}^*(\mathbf{r}) d^3\mathbf{r} = 1 \tag{154}$$

The pair correlation function $g_{ei}^*(\mathbf{r})$ takes account only of the correlations between the electrons and the central ion. In terms of the putative ion pair distribution function $\rho_{ii}(\mathbf{r}) = n_i(1 + g_{ii}(\mathbf{r}))$, which properly accounts for the other ions in the plasma and the correlations between them, the total electron PDF with respect to any ion in the plasma is then given by the convolution, in which the jellium contribution is first subtracted out,

$$\rho_{ei}(\mathbf{r}) = \left(\rho_{ei}^*(\mathbf{r}) - Z^*\rho_0^*(\mathbf{r})\right) + \int \left(\rho_{ei}^*(\mathbf{r}-\mathbf{r}') - Z^*\rho_0^*(\mathbf{r}-\mathbf{r}')\right)\rho_{ii}(\mathbf{r}')d^3\mathbf{r}'$$

$$= Zn_i\left(g_{ei}^*(\mathbf{r}) + n_i\int g_{ei}^*(\mathbf{r}-\mathbf{r}')(1+g_{ii}(\mathbf{r}'))d^3\mathbf{r}'\right) \tag{155}$$

$$\equiv Zn_i(1+g_{ei}(\mathbf{r}))$$

Hence, making use of (154),

$$g_{ei}(\mathbf{r}) = g_{ei}^*(\mathbf{r}) + n_i\int g_{ei}^*(\mathbf{r}-\mathbf{r}')g_{ii}(\mathbf{r}')d^3\mathbf{r}' \tag{156}$$

The Fourier transform of (156) yields

$$g_{ei}(\mathbf{q}) = g_{ei}^*(\mathbf{q})(1+g_{ii}(\mathbf{q})) \tag{157}$$

where the Fourier transforms of the pair correlation functions are defined, in accordance with (111), as follows

$$g_{ei}(\mathbf{q}) = n_i\sqrt{Z}\int g_{ei}(\mathbf{r})e^{-i\mathbf{q}\cdot\mathbf{r}}d^3\mathbf{r}$$

$$g_{ei}^*(\mathbf{q}) = n_i\sqrt{Z}\int g_{ei}^*(\mathbf{r})e^{-i\mathbf{q}\cdot\mathbf{r}}d^3\mathbf{r} \tag{158}$$

$$g_{ii}(\mathbf{q}) = n_i\int g_{ii}(\mathbf{r})e^{-i\mathbf{q}\cdot\mathbf{r}}d^3\mathbf{r}$$

where the integrals span the effectively infinite volume of the system, and exist, free of singularities, in the domain of real $\mathbf{q}$ on account of the existence of finite volume integrals (151). The integral over the ion-ion PCF is generally given by the classical formula [35],

$$n_i\int g_{ii}(\mathbf{r})d^3\mathbf{r} = \frac{\langle\Delta N_i^2\rangle}{\langle N_i\rangle} - 1$$

$$= \frac{n_i T_i}{\kappa_i^T} - 1 \tag{159}$$

where $\langle N_i\rangle = n_i\mathfrak{V}$ and $\langle\Delta N_i^2\rangle$ are the mean and variance of the number of nuclei enclosed within some arbitrary volume, $\mathfrak{V}$, within which the system can be regarded as homogeneous, and $\kappa_i^T$ is the isothermal bulk modulus. The ion-ion pair correlation function thus accounts for the fluctuations in the number of ions within a given volume of plasma. The integral (159) varies between zero, in the weak coupling limit, and -1, in the strong coupling limit. Hence, from (156) and (154),

$$n_i\int g_{ei}(\mathbf{r})d^3\mathbf{r} = \frac{n_i T_i}{\kappa_i^T} \tag{160}$$

which confirms that, on average, the electron fluctuations match the ion fluctuations so as to maintain neutrality. The charge distribution around the ion is, where $n_e = Zn_i$ is the average total electron density,

$$\rho_c(\mathbf{r}) = Zn_i(1 + g_{ii}(\mathbf{r})) - n_e(1 + g_{ei}(\mathbf{r}))$$

$$= Zn_i(g_{ii}(\mathbf{r}) - g_{ei}(\mathbf{r})) \tag{161}$$

the volume integral of which is

$$\int\rho_c(\mathbf{r})d^3\mathbf{r} = Zn_i\int(g_{ii}(\mathbf{r}) - g_{ei}(\mathbf{r}))d^3\mathbf{r}$$

$$= -Z \tag{162}$$

which correctly gives the net excess negative charge correlated with the central ion as that which is necessary to guarantee system neutrality.

Equation (157) is the key result here. It shows that, in terms of $g_{ei}^*(\mathbf{q})$ defined in accordance with (153), which involves only regular well-behaved volume-integrable functions satisfying (151), the total electron-ion pair correlation function takes the same form as for entirely free electrons.

However, one should understand that the determination of the electron-electron correlation function, using the techniques given here, depends upon the validity of (110), which holds for classical and/or weakly interacting particles, but not strictly for bound electrons. The treatment of the strong correlations between closely-coupled electrons within atomic systems is a pervasive problem and is invariably treated via approximations. The fact that methods like DFT, which generally uses an approximated exchange and correlation energy functional, and Hartree-Fock, which treats exchange exactly while ignoring many-body correlations, can be made to give reasonable results is encouragement that the application of such models to bound electrons has at least some limited validity.

### 4.3  Green functions and dielectric function in static local field approximation

The foregoing provides a complete closed-form solution for a many-body Coulomb system that is fully consistent with the given static pair-correlation functions. This is demonstrated below for a system of any number of different charge species.

Substituting for $\Lambda_{ab}(\mathbf{q})$ according to (141), into (49) yields a closed-form equation for the Green function

$$\mathrm{K}_{ab}(\mathbf{q},\omega) = \Pi^0_{aa}(\mathbf{q},\omega)\delta_{ab} + \sqrt{\frac{n_a}{n_b}}\frac{D(\mathbf{q},0)}{D(\mathbf{q},\omega)}\frac{\Pi^0_{aa}(\mathbf{q},\omega)}{\Pi^0_{aa}(\mathbf{q},0)} g'_{ab}(\mathbf{q})\Pi^0_{bb}(\mathbf{q},\omega) \tag{163}$$

which is equivalent to the equations

$$\mathrm{K}_{ab}(\mathbf{q},\omega) = \hat{\mathrm{K}}_{ab}(\mathbf{q},\omega)\Pi^0_{bb}(\mathbf{q},0)$$

$$\hat{\mathrm{K}}_{ab}(\mathbf{q},\omega) = \delta_{ab}\hat{\Pi}^0_{bb}(\mathbf{q},\omega) + \sqrt{\frac{n_a}{n_b}}\frac{1}{\hat{D}(\mathbf{q},\omega)}\hat{\Pi}^0_{aa}(\mathbf{q},\omega) g'_{ab}(\mathbf{q})\hat{\Pi}^0_{bb}(\mathbf{q},\omega) \tag{164}$$

$$\hat{\Pi}^0_{aa}(\mathbf{q},\omega) = \frac{\Pi^0_{aa}(\mathbf{q},\omega)}{\Pi^0_{aa}(\mathbf{q},0)}; \quad \hat{D}(\mathbf{q},\omega) = \frac{D(\mathbf{q},\omega)}{D(\mathbf{q},0)}$$

where $g'_{ab}$ is given by (129). In terms of the above, a combination of equations (52) and (141) yields the frequency-dependent *effective* dielectric function as follows:

$$D(\mathbf{q},\omega) = 1 - D(\mathbf{q},0) \sum_a \hat{\Pi}^0_{aa}(\mathbf{q},\omega) g'_{aa}(\mathbf{q}) \tag{165}$$

where the static function $D(\mathbf{q},0)$ is obtainable from (128). Equations (163) - (165) show that the dynamical effects are essentially those embodied in the RPA, which is unsurprising given that only information about the static correlations has been added. However $D(\mathbf{q},\omega)$ is defined in terms of the effective polarizability, by (52), and is not the true dielectric function; rather it is the RPA dielectric function for the effective potentials $v^{\text{eff}}$ and serves to relate the response function $\mathbf{K}$ to the RPA susceptibilities $\mathbf{\Pi}^0$. Clearly, for $v^{\text{eff}} \neq v$, $\mathbf{\Pi}^0$ and $\mathbf{K}$ do not satisfy the screening equation (15) in terms of the physical Coulomb potentials $v$. To determine the true dielectric function and $\mathbf{\Pi}$, which is the polarization part of $\mathbf{K}$ given by (163), we must return to equation (15) and follow exactly the procedure described in section 2.3, except that we now solve for $\mathbf{\Pi}$ instead of $\mathbf{K}$. In terms of the function $\varepsilon(\mathbf{q},\omega)$ defined by

$$\varepsilon^{-1}(\mathbf{q},\omega) = 1 + \text{trace}(v\mathbf{K}) \tag{166}$$

the matrix $\dfrac{v\mathbf{K}}{\varepsilon^{-1}-1}$ is idempotent, and hence (APPENDIX A, lemma 2) the inverse of $\mathbf{1}+v\mathbf{K}$ is given by

$$(\mathbf{1}+v\mathbf{K})^{-1} = 1 - \varepsilon v\mathbf{K} \tag{167}$$

The required solution of (15) for $\mathbf{\Pi}$ is therefore

$$\mathbf{\Pi} = \mathbf{K} - \varepsilon \mathbf{K} v \mathbf{K} \tag{168}$$

which, again using the idempotency property of $\dfrac{v\mathbf{K}}{\varepsilon^{-1}-1}$, leads to

$$v\mathbf{K} = \frac{v\mathbf{\Pi}}{\varepsilon} \tag{169}$$

which is identical to (26) confirming that $\epsilon(\mathbf{q},\omega)$ defined by (166) is the dielectric function. In terms of $\mathbf{\Pi}$ given by (168), $\varepsilon$ also satisfies

$$\varepsilon(\mathbf{q},\omega) = 1 - \text{trace}(v\mathbf{\Pi}) \tag{170}$$

which confirms that the dielectric function $D(\mathbf{q},\omega)$ as originally defined in section 2.3 is identical to $\varepsilon(\mathbf{q},\omega)$ as defined above. However, with the introduction of the local field corrections in section 3.2, $D(\mathbf{q},\omega)$ became the effective dielectric function ($D(\mathbf{q},\omega) = 1 - \text{trace}(v^{\text{eff}}\mathbf{\Pi}^0)$) being the RPA dielectric function for the LFC modified potential. Thus, having determined an approximation to the full response Green function $\mathbf{K}$ within the static local field approximation involving an effective potential, it is necessary to work back to determine the polarization part $\mathbf{\Pi}$ and the dielectric function $\epsilon(\mathbf{q}, \omega)$ that are related to it via the screening equation involving the physical (Coulomb) potentials. In this way, the method provides a fully self-consistent means of incorporating the short-range static correlations into the RPA.

## 5  CONCLUSIONS

A formulation of a quantum-mechanical model of the electron-ion temperature relaxation in Coulomb plasmas, which includes effects of plasma non-ideality through static local field corrections, is presented. The method uses a first principles approach in which a weak coupling approximation is used only in respect of the cross-coupling between different species, which is entirely consistent with the notion that temperature separation between these species should be possible. The approximation consists of neglecting the off-diagonal (with respect to species labels) terms in the susceptibility matrix. Such terms are already vanishing in the Random Phase Approximation, for which the elements $\Pi_{aa}$ of the susceptibility are given, for each species, by the free-particle Lindhard formula. The RPA therefore provides a convenient basis for the model to which the short-range static correlations between all the various particle types are restored through the use of local field corrections.

The general result for the two component plasma given by (72) is similar to one given recently by Daligault and Dimonte [12].

A practical form of the result for the electron-ion energy exchange coefficient in a hot plasma (temperatures $\gg \Omega_i$) is basically given by (85). Equation (87) is a new result, and incorporates, in addition to the local field correction for the electron-ion correlations, plasma quantum effects associated with ion-acoustic modes and a treatment of quantum diffraction applicable to all regimes of electron degeneracy. Dynamical screening is accounted for, though the approximation made at equation (80) means that the screening of heavy particles by electrons is effectively static.

Imposing the requirement that all plasma components interact solely via Coulomb potentials, ie employing a physical all-electron picture without pseudopotentials, leads to the simplifying relations expressed by (54) and (131)-(132) which mean that all the pair correlation functions, and hence the local field corrections, for a two-component system, can be obtained from knowledge of only two of them. Note that, for subsystems that are not in thermal equilibrium, ie $T_a \neq T_b$, some matrices, such as the provisional pair correlation function $\tilde{g}_{ab}$ that satisfies the generalized quantum Ornstein-Zernicke relation, are not symmetric. Therefore, from a classical determination of the ion-ion correlation function, $g_{ii}$, and the electron ion correlation function $g_{ei}^*$ given by a single-centre average-atom calculation, for example, it is possible to determine all the components of the pair correlation matrix of a two-component plasma, and hence the static LFCs.

The model includes the quantum exchange corrections in a consistent manner through the RPA exchange correlation and susceptibility functions, $g_{aa}^0(\mathbf{q})$ and $\mathbf{\Pi}_{aa}^0(\mathbf{q},0)$ respectively, and leads ultimately to a simple closed formula for the Green functions $K_{ab}(\mathbf{q},\omega)$, given by equations (163) - (164) from which a fully self-consistent approximate solution of the Coulomb many body problem ensues, in which the static correlations are, in principle, treated exactly. In this respect, the model can be considered to represent a significant improvement over pure RPA.

However the model is approximate in that fails to treat short-range dynamical correlations, which may be associated with processes such a non-radiative collisional ionization and recombination. Also, by assuming that the only interactions in the system occur via static Coulomb interactions, no account is taken of non-Coulomb dynamical interactions, such as one-electron radiative processes occurring in the nuclear field. The model does not therefore incorporate a description of the high-frequency radiative opacity and it should, in principle, be possible to include atomic photoabsorption via dynamical local field corrections. However Bremsstrahlung is a particular photoabsorption process that is intrinsically linked to the dynamic fluctuations involving electron-ion correlations, which are typically treated as short-range "collisions" between free electrons and ions. It is a general, if not reasonable, assumption that these can be treated as a small perturbation affecting the state of the plasma. That is to say, while the state of plasma has a determining effect on the bremsstrahlung, the bremsstrahlung has an insignificant effect on the plasma state. However existing models typically treat the coupling between radiation and plasma as an entirely separate process thus neglecting potentially important correlations between different processes that are treated as pairwise interactions (electron-ion, electron-radiation). Bremsstrahlung, for example, simultaneously involves the interactions between electrons, ions and radiation. A fully unified treatment of electron-ion bremsstrahlung and plasma fluctuations would be a subject for further study.

So, while, in common with the RPA on which it is based, the model clearly does not treat high frequency dynamic effects, such as photoabsorption, it would be expected to provide an enhanced description of the static and low-frequency properties of a Coulomb plasma or simple metallic liquid, including equation of state and non-radiative transport properties.

Density fluctuations and temperature relaxation…	45## 6 ACKNOWLEDGEMENTS

The author would like to acknowledge useful discussions with Gianluca Gregori and Lee Pattison on some of the issues arising in this paper, and for their helpful suggestions.

# APPENDIX A  MATHEMATICAL THEOREMS AND LEMMAS

## A.1	Some theorems concerning matrices

Lemma 1

If $\mathbf{P}$ is an idempotent matrix ($\mathbf{P}^2 = \mathbf{P}$) then the determinant $\text{Det}(\mathbf{1}+\lambda\mathbf{P}) = |\mathbf{1}+\lambda\mathbf{P}|$ is given by

$$|\mathbf{1}+\lambda\mathbf{P}| = 1+\lambda \tag{171}$$

[Proof: Let $D(\lambda) = |\mathbf{1}+\lambda\mathbf{P}|$. By induction, for any set of numbers $\{\lambda_i\}$,

$$\prod_i (\mathbf{1}+\lambda_i\mathbf{P}) = \mathbf{1} + \left\{\prod_i (1+\lambda_i) - 1\right\}\mathbf{P} \tag{172}$$

Taking the determinant of both sides of (172) yields

$$\prod_i D(\lambda_i) = D\left(\prod_i (1+\lambda_i) - 1\right) \tag{173}$$

Which holds for any set of values $\{\lambda_i\}$, if and only if

$$D(\lambda) = 1+\lambda \tag{174}$$

QED]

Lemma 2

If $\mathbf{P}$ is an idempotent matrix ($\mathbf{P}^2 = \mathbf{P}$) then the inverse of the matrix $\mathbf{1}+\lambda\mathbf{P}$ for $\lambda \neq -1$ is given by

$$(\mathbf{1}+\lambda\mathbf{P})^{-1} = \mathbf{1} - \frac{\lambda}{1+\lambda}\mathbf{P} \tag{175}$$

[Proof: By lemma 1, if $\lambda \neq -1$, the matrix $\mathbf{1}+\lambda\mathbf{P}$ has a non-vanishing determinant and therefore a unique inverse.

Multiplying the right hand side of (175) by $\mathbf{1}+\lambda\mathbf{P}$ yields,

$$(\mathbf{1}+\lambda\mathbf{P})\left(\mathbf{1} - \frac{\lambda}{1+\lambda}\mathbf{P}\right) = \left(\mathbf{1} - \frac{\lambda}{1+\lambda}\mathbf{P}\right)(\mathbf{1}+\lambda\mathbf{P}) = \mathbf{1} + \lambda\mathbf{P} - \frac{\lambda}{1+\lambda}\mathbf{P} - \frac{\lambda^2}{1+\lambda}\mathbf{P} = \mathbf{1} \tag{176}$$

QED]

## A.2     Properties of the density correlation matrix for a multicomponent plasma

This appendix section proves the properties of the correlation matrix $S_{ab}(\mathbf{q},\omega)$ defined by (1).

### A.2.1     Hermiticity

Taking the complex conjugate of (1), using the fact that $\langle \delta\hat{n}_b(\mathbf{r}',t')\delta\hat{n}_a(\mathbf{r},t)\rangle$ is a function only of $\mathbf{r}-\mathbf{r}'$ and $t-t'$,

$$S_{ab}^*(\mathbf{q},\omega) = \frac{1}{2\pi n_e \mathfrak{V}} \iint_{\mathfrak{V}} d^3\mathbf{r}\, d^3\mathbf{r}' \int_{-\infty}^{+\infty} dt\, \langle \delta\hat{n}_a(\mathbf{r},t)\delta\hat{n}_b(\mathbf{r}',t')\rangle^* e^{i\mathbf{q}\cdot(\mathbf{r}-\mathbf{r}')} e^{-i\omega(t-t')}$$

$$= \frac{1}{2\pi n_e \mathfrak{V}} \iint_{\mathfrak{V}} d^3\mathbf{r}\, d^3\mathbf{r}' \int_{-\infty}^{+\infty} dt\, \langle \delta\hat{n}_b(\mathbf{r}',t')\delta\hat{n}_a(\mathbf{r},t)\rangle e^{i\mathbf{q}\cdot(\mathbf{r}-\mathbf{r}')} e^{-i\omega(t-t')}$$

$$= \frac{1}{2\pi n_e \mathfrak{V}} \iint_{\mathfrak{V}} d^3\mathbf{r}\, d^3\mathbf{r}' \int_{-\infty}^{+\infty} dt'\, \langle \delta\hat{n}_b(\mathbf{r},t)\delta\hat{n}_a(\mathbf{r}',t')\rangle e^{-i\mathbf{q}\cdot(\mathbf{r}-\mathbf{r}')} e^{i\omega(t-t')} \qquad (177)$$

$$= \frac{1}{2\pi n_e \mathfrak{V}} \iint_{\mathfrak{V}} d^3\mathbf{r}\, d^3\mathbf{r}' \int_{-\infty}^{+\infty} dt\, \langle \delta\hat{n}_b(\mathbf{r},t)\delta\hat{n}_a(\mathbf{r}',t')\rangle e^{-i\mathbf{q}\cdot(\mathbf{r}-\mathbf{r}')} e^{i\omega(t-t')}$$

$$= S_{ba}(\mathbf{q},\omega)$$

QED]

## A.2.2 Time reversality

Using the time reversal property of the expectation value of the product of density operator,

$$\langle \delta\hat{n}_b(\mathbf{r}',t')\delta\hat{n}_a(\mathbf{r},t)\rangle = \langle \delta\hat{n}_b(\mathbf{r},-t)\delta\hat{n}_a(\mathbf{r}',-t')\rangle \tag{178}$$

yields

$$S_{ab}(-\mathbf{q},\omega) = \frac{1}{2\pi n_e \mathfrak{V}} \iint_{\mathfrak{V}} d^3\mathbf{r}\, d^3\mathbf{r}' \int_{-\infty}^{+\infty} dt \langle \delta\hat{n}_a(\mathbf{r},t)\delta\hat{n}_b(\mathbf{r}',t')\rangle e^{i\mathbf{q}\cdot(\mathbf{r}-\mathbf{r}')} e^{i\omega(t-t')}$$

$$= \frac{1}{2\pi n_e \mathfrak{V}} \iint_{\mathfrak{V}} d^3\mathbf{r}\, d^3\mathbf{r}' \int_{-\infty}^{+\infty} dt' \langle \delta\hat{n}_a(\mathbf{r}',t')\delta\hat{n}_b(\mathbf{r},t)\rangle e^{-i\mathbf{q}\cdot(\mathbf{r}-\mathbf{r}')} e^{-i\omega(t-t')}$$

$$= \frac{1}{2\pi n_e \mathfrak{V}} \iint_{\mathfrak{V}} d^3\mathbf{r}\, d^3\mathbf{r}' \int_{-\infty}^{+\infty} dt' \langle \delta\hat{n}_a(\mathbf{r},-t)\delta\hat{n}_b(\mathbf{r}',-t')\rangle e^{-i\mathbf{q}\cdot(\mathbf{r}-\mathbf{r}')} e^{-i\omega(t-t')} \tag{179}$$

$$= \frac{1}{2\pi n_e \mathfrak{V}} \iint_{\mathfrak{V}} d^3\mathbf{r}\, d^3\mathbf{r}' \int_{-\infty}^{+\infty} dt \langle \delta\hat{n}_a(\mathbf{r},t)\delta\hat{n}_b(\mathbf{r}',t')\rangle e^{-i\mathbf{q}\cdot(\mathbf{r}-\mathbf{r}')} e^{i\omega(t-t')}$$

$$= S_{ab}(\mathbf{q},\omega)$$

QED]

### A.2.3 KMS relation

This is obtained by expanding the expectation value in terms of the trace of the thermally-weighted operator matrix,

$$S_{ab}(\mathbf{q},\omega) = \frac{1}{2\pi n_e \mathfrak{V}} \int_{-\infty}^{+\infty} \langle \delta\hat{n}_a(\mathbf{q},t)\delta\hat{n}_b(-\mathbf{q},t') \rangle e^{i\omega(t-t')} dt$$

$$= \frac{1}{2\pi n_e \mathfrak{V}} \sum_{\alpha,\beta} \int_{-\infty}^{+\infty} \langle \alpha|\delta\hat{n}_a(\mathbf{q},t)|\beta\rangle\langle\beta|\delta\hat{n}_b(-\mathbf{q},t')e^{-\hat{H}/T}|\alpha\rangle e^{i\omega(t-t')} dt$$

$$= \frac{1}{n_e \mathfrak{V}} \sum_{\alpha,\beta} e^{-\varepsilon_\alpha/T} \langle\alpha|\delta\hat{n}_a(\mathbf{q},0)|\beta\rangle\langle\beta|\delta\hat{n}_b(-\mathbf{q},0)|\alpha\rangle \delta(\varepsilon_\alpha - \varepsilon_\beta + \omega)$$

$$= \frac{1}{n_e \mathfrak{V}} \sum_{\alpha,\beta} e^{-\varepsilon_\beta/T} \langle\beta|\delta\hat{n}_a(\mathbf{q},0)|\alpha\rangle\langle\alpha|\delta\hat{n}_b(-\mathbf{q},0)|\beta\rangle \delta(\varepsilon_\alpha - \varepsilon_\beta - \omega)$$

$$= \frac{e^{\omega/T}}{n_e \mathfrak{V}} \sum_{\alpha,\beta} e^{-\varepsilon_\alpha/T} \langle\alpha|\delta\hat{n}_b(-\mathbf{q},0)|\beta\rangle\langle\beta|\delta\hat{n}_a(\mathbf{q},0)|\alpha\rangle \delta(\varepsilon_\alpha - \varepsilon_\beta - \omega)$$

$$= e^{\omega/T} S_{ba}(-\mathbf{q},-\omega)$$

$$= e^{\omega/T} S_{ba}(\mathbf{q},-\omega) \tag{180}$$

## APPENDIX B  DERIVATION OF THE FORMULA (EQUATION (110)) FOR THE STATIC PAIR DISTRIBUTION FUNCTION

The static pair distribution function $g(\mathbf{q})$ is given, for $T = T_a = T_b$, in terms of the dynamic structure factor $S(\mathbf{q}, \omega)$, as defined by (1), by the following standard relations,

$$g_{ab}(\mathbf{q}) + \delta_{ab} = \frac{n_e}{\sqrt{n_a n_b}} \hat{S}_{ab}(\mathbf{q})$$

$$= \frac{n_e}{\sqrt{n_a n_b}} \int_{-\infty}^{+\infty} \hat{S}(\mathbf{q}, \omega) \, d\omega$$

(181)

which, using the Fluctuation Dissipation Theorem (36), becomes

$$g_{ab}(\mathbf{q}) + \delta_{ab} = \frac{-1}{2\pi \sqrt{n_a n_b}} \int_{-\infty}^{+\infty} \coth\left(\frac{\omega}{2T}\right) K''_{ab}(\mathbf{q}, \omega) \, d\omega$$

$$= \frac{-1}{\pi \sqrt{n_a n_b}} \int_{0}^{+\infty} \coth\left(\frac{\omega}{2T}\right) \operatorname{Re} K''_{ab}(\mathbf{q}, \omega) \, d\omega$$

(182)

using the fact noted in sect. 2.2 that the real and imaginary parts of $K''_{ab}(\mathbf{q}, \omega)$ are respectively odd and even functions of $\omega$. We now make use of the spectral representation [16], [28],

$$\mathbf{K}(\mathbf{q}, \omega) = \frac{1}{2\pi} \int_{-\infty}^{+\infty} \frac{\boldsymbol{\sigma}(\mathbf{q}, \omega')}{\omega - \omega' + i 0^+} \, d\omega'$$

(183)

where $\boldsymbol{\sigma}(\mathbf{q}, \omega) = \|\sigma_{ab}(\mathbf{q}, \omega)\|$ is the spectral density, which, in this context, takes the form of a Hermitian matrix with the following properties on $\operatorname{Im} \omega = 0$,

$$\operatorname{Re} \boldsymbol{\sigma}(\mathbf{q}, -\omega) = -\operatorname{Re} \boldsymbol{\sigma}(\mathbf{q}, \omega)$$

$$\operatorname{Im} \boldsymbol{\sigma}(\mathbf{q}, -\omega) = \operatorname{Im} \boldsymbol{\sigma}(\mathbf{q}, \omega)$$

$$\boldsymbol{\sigma}(-\mathbf{q}, \omega) = \boldsymbol{\sigma}(\mathbf{q}, \omega)$$

$$\lim_{\omega \to \pm\infty} \left(\omega^2 \boldsymbol{\sigma}(\mathbf{q}, \omega)\right) = \mathbf{0}$$

(184)

$$\omega \operatorname{Re} \sigma_{aa}(\mathbf{q}, \omega) \geq 0 \tag{185}$$

$$\int_0^{+\infty} \operatorname{Re} \sigma_{ab}(\mathbf{q}, \omega) \omega \, d\omega = \pi \delta_{ab} \frac{n_a q^2}{m_a} \tag{186}$$

$$\int_0^{\infty} \boldsymbol{\sigma}(\mathbf{q}, \omega) \frac{d\omega}{\omega} = -\pi \mathbf{K}(\mathbf{q}, 0) \tag{187}$$

and in terms of which

$$\mathbf{K}''(\mathbf{q}, \omega) = -\tfrac{1}{2} \boldsymbol{\sigma}(\mathbf{q}, \omega) \tag{188}$$

whereupon (182) becomes

$$g_{ab}(\mathbf{q}) + \delta_{ab} = \frac{1}{2\pi \sqrt{n_a n_b}} \int_0^{+\infty} \coth\left(\frac{\omega}{2T}\right) \operatorname{Re} \sigma_{ab}(\mathbf{q}, \omega) \, d\omega$$

$$= \frac{T}{\pi \sqrt{n_a n_b}} \int_0^{+\infty} \left(\frac{\omega}{2T}\right) \coth\left(\frac{\omega}{2T}\right) \operatorname{Re} \sigma_{ab}(\mathbf{q}, \omega) \frac{d\omega}{\omega} \tag{189}$$

Equation (185) is a necessary condition for the system to exhibit a stable response to small disturbances about equilibrium, and equation (186) incorporates the f-sum rule. Equation (187) follows directly from (183).

Now, using the expansion, $x \coth(x) = 1 + \tfrac{1}{3} x^2 + \mathcal{O}(x^4)$, together with (186) and (187), this can be written in the form

$$g_{ab}(\mathbf{q}) = \frac{T}{\pi \sqrt{n_a n_b}} \int_0^{+\infty} \left(\left(\frac{\omega}{2T}\right) \coth\left(\frac{\omega}{2T}\right) - 1 - \frac{\omega^2}{12 T^2}\right) \operatorname{Re} \sigma_{ab}(\mathbf{q}, \omega) \frac{d\omega}{\omega}$$

$$- \frac{T}{\sqrt{n_a n_b}} \mathbf{K}_{ab}(\mathbf{q}, 0) - \delta_{ab} \left(1 - \frac{q^2}{12 m_a T}\right) \tag{190}$$

In the case of non-interacting particles, this becomes

$$g_{ab}^0(\mathbf{q}) = \delta_{ab} g_{aa}^0(\mathbf{q})$$

$$g_{aa}^0(\mathbf{q}) = \frac{T}{\pi n_a} \int_0^{+\infty} \left(\left(\frac{\omega}{2T}\right) \coth\left(\frac{\omega}{2T}\right) - 1 - \frac{\omega^2}{12 T^2}\right) \operatorname{Re} \sigma_{aa}^0(\mathbf{q}, \omega) \frac{d\omega}{\omega} \tag{191}$$

$$- \frac{T}{n_a} \Pi_{aa}^0(\mathbf{q}, 0) - 1 + \frac{q^2}{12 m_a T}$$

Subtracting (191) from (190) then yields

$$g_{ab}(\mathbf{q}) - g_{ab}^0(\mathbf{q}) = \frac{T}{\pi\sqrt{n_a n_b}} \int_0^{+\infty} \left(\left(\frac{\omega}{2T}\right)\coth\left(\frac{\omega}{2T}\right) - 1 - \frac{\omega^2}{12T^2}\right) \left(\operatorname{Re}\sigma_{ab}(\mathbf{q},\omega) - \delta_{ab}\sigma_{aa}^0(\mathbf{q},\omega)\right) \frac{d\omega}{\omega}$$

(192)

$$-\frac{T}{\sqrt{n_a n_b}}\left(\mathrm{K}_{ab}(\mathbf{q},0) - \delta_{ab}\Pi_{aa}^0(\mathbf{q},0)\right)$$

Thus far the calculation is exact. We now make use of the fact that, for large excitations, the interactions between particles become unimportant and the particles become effectively non-interacting, so that, as $\omega \to \infty$, $\operatorname{Re}\sigma_{ab}(\mathbf{q},\omega) \sim \delta_{ab}\sigma_{aa}^0(\mathbf{q},\omega)$. Specifically, we make a semiclassical approximation by assuming that $\sigma_{aa}(\mathbf{q},\omega) \sim \sigma_{aa}^0(\mathbf{q},\omega)$ and $\operatorname{Re}\sigma_{ab}(\mathbf{q},\omega) \sim 0$, $a \neq b$, for $\omega \gtrsim T$. This yields

$$g_{ab}(\mathbf{q}) - g_{ab}^0(\mathbf{q}) \simeq -\frac{T}{\sqrt{n_a n_b}}\left(\mathrm{K}_{ab}(\mathbf{q},0) - \delta_{ab}\Pi_{aa}^0(\mathbf{q},0)\right)$$

(193)

which, when combined with the defining equation (49) for the effective potential, leads to

$$g_{ab}(\mathbf{q}) - g_{ab}^0(\mathbf{q}) \simeq -\frac{T}{\sqrt{n_a n_b}} \Pi_{aa}^0(\mathbf{q},0) \sum_c v_{ac}^{\mathrm{eff}}(\mathbf{q}) \mathrm{K}_{cb}(\mathbf{q},0)$$

(194)

$$= -\frac{T}{\sqrt{n_a n_b}} \Pi_{bb}^0(\mathbf{q},0) \sum_c \mathrm{K}_{ac}(\mathbf{q},0) v_{cb}^{\mathrm{eff}}(\mathbf{q})$$

making use of the fact that $\mathbf{K}(\mathbf{q},0)$ is real and symmetric and that $g_{ab}(\mathbf{q})$ and $\Pi^0(\mathbf{q},0)$ are real. Finally, noting that the assumption leading to (193) holds only for non-degenerate (classical) systems, we use the classical limits $\Pi_{bb}^0(\mathbf{q},0) \sim -\frac{n_b}{T}$ and $g_{ab}^0(\mathbf{q}) \simeq 0$ to obtain

$$g_{ab}(\mathbf{q}) \simeq \sqrt{\frac{n_b}{n_a}} \sum_c \mathrm{K}_{ac}(\mathbf{q},0) v_{cb}^{\mathrm{eff}}(\mathbf{q})$$

(195)

Equation (195) is the classical form of the result for systems in thermal equilibrium, and demonstrates the appropriateness of using the effective potential $v^{\mathrm{eff}}(\mathbf{q})$ here, rather than $v(\mathbf{q})$.

The right hand side represents the response of the density of particles $a$ to the potential due to particle $b$, as if that potential were imposed as an external force. Quantum mechanically however, since $b$ is actually a particle in the plasma, there is also a response purely due to exchange correlations between the particle and the rest of the plasma. In particular, in the limit of vanishing potential interaction, $v_{ab}(\mathbf{q}) \to 0$, $v_{ab}^{\text{eff}}(\mathbf{q}) \to 0$, $\forall a$, between the source particle $b$ and the rest of the plasma, $g_{ab}(\mathbf{q}) \to g_{ab}^0(\mathbf{q})$.

We first consider the case of a single particle species, $a = b$. In terms of

$$I \equiv \frac{T_a}{\pi n_a} \int_0^{+\infty} \left(\frac{\omega}{2T_a}\right) \coth\left(\frac{\omega}{2T_a}\right) \operatorname{Re} \sigma_{aa}(\mathbf{q},\omega) \frac{d\omega}{\omega} = 1 + g_{aa}(\mathbf{q}) \tag{196}$$

$$J \equiv \frac{T_a}{\pi n_a} \int_0^{+\infty} \operatorname{Re} \sigma_{aa}(\mathbf{q},\omega) \frac{d\omega}{\omega} = -\frac{T_a}{n_a} K_{aa}(\mathbf{q},0) \tag{197}$$

using (185), and the property $1 + \tfrac{1}{3} x^2 \geq x \coth(x) \geq 1$, together with the f-sum rule (186), the following inequalities can be readily obtained

$$J + \tfrac{1}{3} s \geq I \geq J \geq I - \tfrac{1}{3} s \tag{198}$$

where $s = q^2/4 m_a T_a$, while, for fermions

$$1 \geq I \geq J \geq 0 \tag{199}$$

In particular, it follows from (198) that $I \to J$ as $q \to 0$. In addition, by application of the Cauchy-Schwarz inequality to various combinations of the integrals,

$$J^0 I \geq K^2$$
$$I^0 J \geq K^2 \tag{200}$$

$$sJ \geq L^2, \quad sJ^0 \geq (L^0)^2 \tag{201}$$

where

$$I^0 \equiv \frac{T_a}{\pi n_a} \int_0^{+\infty} \left(\frac{\omega}{2T_a}\right) \coth\left(\frac{\omega}{2T_a}\right) \operatorname{Re} \sigma_{aa}^0(\mathbf{q},\omega) \frac{d\omega}{\omega} = 1 + g_{aa}^0(\mathbf{q}) \equiv R_a(\mathbf{q}) \tag{202}$$

$$J^0 \equiv \frac{T_a}{\pi n_a} \int_0^{+\infty} \operatorname{Re} \sigma_{aa}^0(\mathbf{q},\omega) \frac{d\omega}{\omega} = -\frac{T_a}{n_a} \Pi_{aa}^0(\mathbf{q},0) \equiv \tilde{R}_a(\mathbf{q}) \tag{203}$$

$$K \equiv \frac{T_a}{\pi n_a} \int_0^{+\infty} \left(\left(\frac{\omega}{2T_a}\right) \coth\left(\frac{\omega}{2T_a}\right) \operatorname{Re} \sigma_{aa}(\mathbf{q},\omega) \operatorname{Re} \sigma_{aa}^0(\mathbf{q},\omega)\right)^{1/2} \frac{d\omega}{\omega} \tag{204}$$

$$L \equiv \frac{1}{2\pi n_a} \int_0^{+\infty} \operatorname{Re} \sigma_{aa}(\mathbf{q},\omega) d\omega, \quad L^0 \equiv \frac{1}{2\pi n_a} \int_0^{+\infty} \operatorname{Re} \sigma_{aa}^0(\mathbf{q},\omega) d\omega, \tag{205}$$

where $R_a(\mathbf{q})$ and $\tilde{R}_a(\mathbf{q})$ are identical to the functions $R_\mathbf{q}^a$ and $\tilde{R}_\mathbf{q}^a$ defined in ref.[28] The products $J^0 I$ and $I^0 J$ are therefore subject to similar bounds, and are equal in the limit of $q \to 0$. For large values of $q$, the integrals $I$ and $I_0$ both tend to unity while $J$ and $J_0$ tend to zero as $\mathcal{O}(q^{-2})$. For large $q$, $q^2/2m_a$ ultimately becomes the dominant energy scale, and both $\sigma_{aa}(\mathbf{q},\omega)$ and $\sigma^0_{aa}(\mathbf{q},\omega)$ must then take the form

$$\sigma_{aa}(\mathbf{q},\omega) \sim \sigma^0_{aa}(\mathbf{q},\omega) \sim \frac{2\pi n_a m_a}{q^2} f\left(\frac{2m_a \omega}{q^2}\right) \tag{206}$$

where, in order to comply with the f-sum rule,

$$\int_0^\infty f(x) x \, dx = 1 \tag{207}$$

The basis for this assertion is that, for fermions at least, at close enough distances (large $\mathbf{q}$) exchange always dominates. (Consider that the energy density of a non-interacting Fermi gas increases $\propto n^{5/3}$, whereas the Coulomb energy density of a one-component plasma, in the strong coupling limit, increases $\propto n^{4/3}$.)

The integrals $\int_0^\infty f(x) \, dx$ and $\int_0^\infty f(x) \frac{dx}{x}$ are presumed to exist, in which case they are positive definite numbers satisfying, in accordance with the Cauchy-Schwarz inequality in conjunction with (207),

$$\int_0^\infty f(x) \frac{dx}{x} \geq \left(\int_0^\infty f(x) \, dx\right)^2 \tag{208}$$

which corresponds to (201). Clearly, this limit also yields $J^0 I \simeq I^0 J$. It is now a matter of reasonable conjecture that this equality is at least approximately true *for all* $\mathbf{q}$. This translates to

$$(1 + g_{aa}(\mathbf{q}))\Pi^0_{aa}(\mathbf{q},0) \simeq (1 + g^0_{aa}(\mathbf{q})) K_{aa}(\mathbf{q},0) \tag{209}$$

leading to

$$g_{aa}(\mathbf{q}) \simeq g^0_{aa}(\mathbf{q}) + (1 + g^0_{aa}(\mathbf{q})) K_{aa}(\mathbf{q},0) v^{\text{eff}}_{aa}(\mathbf{q}) \tag{210}$$

which result is entirely consistent with the classical formula (195), to which it reduces when $g^0_{aa}(\mathbf{q}) = 0$.

In generalizing from this to mixtures of dissimilar particles, we take note that the factor of $(1 + g^0_{aa}(\mathbf{q}))$ in the second term of the right hand side of (210) arises because of the exchange correlations involving the source particle. The generalization is therefore postulated to be

$$\tilde{g}_{ab}(\mathbf{q}) \simeq g^0_{ab}(\mathbf{q}) + (1 + g^0_{bb}(\mathbf{q})) \sqrt{\frac{n_b}{n_a}} \sum_c K_{ac}(\mathbf{q},0) v^{\text{eff}}_{cb}(\mathbf{q}) \tag{211}$$

which gives the result as expressed by (111), from which (110) follows directly. However, the function $\tilde{g}_{ab}(\mathbf{q})$ defined by (211) is not, in general, symmetric under $a \leftrightarrow b$, which indicates that

(211) is incomplete. However we shall proceed on the basis that the actual pair correlation function $g_{ab}(\mathbf{q})$, which is necessarily symmetric according to the definition (181), will be an average of $\tilde{g}_{ab}(\mathbf{q})$ and $\tilde{g}_{ba}(\mathbf{q})$.

Using (49), (211) can be written

$$\tilde{g}_{ab}(\mathbf{q}) \simeq \left(1 + g_{bb}^0(\mathbf{q})\right) \sqrt{\frac{n_b}{n_a}} \frac{K_{ab}(\mathbf{q},0)}{\Pi_{bb}^0(\mathbf{q},0)} - \delta_{ab} \tag{212}$$

which, using (49), (134) and (64), becomes

$$\tilde{g}_{ab}(\mathbf{q}) \simeq g_{ab}^0(\mathbf{q}) + \frac{v(\mathbf{q})}{D(\mathbf{q},0)} \sqrt{\frac{n_b}{n_a}} \Pi_{aa}^0(\mathbf{q},0) Q_a Q_b \Lambda_{ab}(\mathbf{q}) \left(1 + g_{bb}^0(\mathbf{q})\right)$$

$$= \delta_{ab}\left(R_b(\mathbf{q}) - 1\right) - \frac{Q_b}{Q_a} \sqrt{\frac{n_b}{n_a}} \frac{1}{D(\mathbf{q},0)} \left(\varepsilon_a^0(\mathbf{q},0) - 1\right) \Lambda_{ab}(\mathbf{q}) R_b(\mathbf{q})$$

$$= \delta_{ab}\left(R_b(\mathbf{q}) - 1\right) - \sqrt{\frac{n_b}{n_a}} \frac{Q_a \Lambda_{ab}(\mathbf{q}) Q_b}{Q_a \Lambda_{aa}(\mathbf{q}) Q_a} \frac{\left(D_a(\mathbf{q},0) - 1\right)}{D(\mathbf{q},0)} R_b(\mathbf{q}) \tag{213}$$

$$= R_b(\mathbf{q}) \frac{Q_a \Lambda_{ab}(\mathbf{q}) Q_b}{Q_a \Lambda_{aa}(\mathbf{q}) Q_a} \sqrt{\frac{n_b}{n_a}} \left(\delta_{ab} - \frac{D_a(\mathbf{q},0) - 1}{D(\mathbf{q},0)}\right) - \delta_{ab}$$

where $D_a(\mathbf{q},\omega) = 1 + \Lambda_{aa}(\mathbf{q})\left(\varepsilon_a^0(\mathbf{q},\omega) - 1\right)$. Support for the postulate (211) is provided by agreement with one of the asymmetric non-collective formulae given in section 4.2.4 of ref.[28] when $\Lambda_{ab} \equiv 1$. The ratio $\tilde{g}_{ab}/\tilde{g}_{ba}$ is yielded as

$$\frac{\tilde{g}_{ab}(\mathbf{q})}{\tilde{g}_{ba}(\mathbf{q})} = \frac{T_b}{T_a} \frac{\tilde{R}_a(\mathbf{q}) R_b(\mathbf{q})}{R_a(\mathbf{q}) \tilde{R}_b(\mathbf{q})} \tag{214}$$

To create a symmetric pair correlation function, we follow previous authors [36], [37] and use the concept of the effective relative temperature, which, for a classical Boltzmann gas, would be defined by

$$T_{ab}^{\text{rel}} = \left(\frac{T_a}{m_a} + \frac{T_b}{m_b}\right)\left(\frac{1}{m_a} + \frac{1}{m_b}\right)^{-1}$$

$$= \frac{m_b T_a + m_a T_b}{m_a + m_b} \tag{215}$$

which is proportional to the product of the mean square relative velocity multiplied by the two particle reduced mass. For fermion particles, it is useful to define effective masses $M_a(\mathbf{q})$ and temperatures $T_a^{\text{eff}}(\mathbf{q})$ by

$$M_a(\mathbf{q}) = m_a \tilde{R}_a(\mathbf{q})$$

$$T_a^{\text{eff}}(\mathbf{q}) = \frac{R_a(\mathbf{q})}{\tilde{R}_a(\mathbf{q})} T_a \quad (216)$$

$$\simeq T_a + \frac{q^2}{12 M_a(\mathbf{q})}$$

A measure of the mean square velocity is then given by $T_a^{\text{eff}}(\mathbf{q})/M_a(\mathbf{q})$, which leads to the generalized effective relative temperature,

$$T_{ab}^{\text{rel}}(\mathbf{q}) = T_{ba}^{\text{rel}}(\mathbf{q}) = \frac{M_b(\mathbf{q}) T_a^{\text{eff}}(\mathbf{q}) + M_a(\mathbf{q}) T_b^{\text{eff}}(\mathbf{q})}{M_b(\mathbf{q}) + M_a(\mathbf{q})} \quad (217)$$

Applying the argument that the pair correlation function must depend upon only the relative velocities, a symmetric pair correlation function may now be defined by

$$g_{ab}(\mathbf{q}) = \tilde{g}_{ab}(\mathbf{q}) \frac{T_a^{\text{eff}}(\mathbf{q})}{T_{ab}^{\text{rel}}(\mathbf{q})}$$

$$= \tilde{g}_{ba}(\mathbf{q}) \frac{T_b^{\text{eff}}(\mathbf{q})}{T_{ab}^{\text{rel}}(\mathbf{q})} \quad (218)$$

which is equivalent to the weighted harmonic average

$$\frac{1}{g_{ab}(\mathbf{q})} = \frac{1}{M_b(\mathbf{q}) + M_a(\mathbf{q})} \left( \frac{M_b(\mathbf{q})}{\tilde{g}_{ab}(\mathbf{q})} + \frac{M_a(\mathbf{q})}{\tilde{g}_{ba}(\mathbf{q})} \right) \quad (219)$$

A situation of particular interest and importance arises when particle $a$ say is much lighter than particle $b$ so that $M_b(\mathbf{q}) \gg M_a(\mathbf{q})$ $\forall \mathbf{q}$, then, provided that $M_b(\mathbf{q}) T_a^{\text{eff}}(\mathbf{q}) \gg M_a(\mathbf{q}) T_b^{\text{eff}}(\mathbf{q})$, this yields $T_{ab}^{\text{rel}}(\mathbf{q}) = T_{ba}^{\text{rel}}(\mathbf{q}) \simeq T_a^{\text{eff}}(\mathbf{q})$ and $g_{ab}(\mathbf{q}) = g_{ba}(\mathbf{q}) \simeq \tilde{g}_{ab}(\mathbf{q})$. This is typically applicable to electrons and ions in a plasma ($a \to \text{e}$, $b \to \text{i}$).

## APPENDIX C LIST OF SYMBOLS

### C.1 List of symbols used for mathematical and physical quantities

Throughout this article, Planck's constant, $\hbar$, and Boltzmann's constant, $k_B$, are consistently set equal to unity, which effectively puts energy, temperature and frequency into the same units.

$a, b, c...$ labels denoting conserved particle species (electrons and atomic nuclei) present in the system.

$\hat{a}(\phi, t)$ Field operator corresponding to annihilation of a single particle in the state $\phi$ at time $t$.

$\hat{a}^\dagger(\phi, t)$ Field operator corresponding to creation of a single particle in the state $\phi$ at time $t$.

$\hat{c}_{ab}$ Symmetrised density-correlation operator as defined by equation (32).

$\mathbf{c}$ $= \|c_{ab}(\mathbf{q})\|$, Direct correlation matrix defined by (121).

$D$ $= D(\mathbf{q}, \omega)$, dielectric function initially defined by (26) and given by $D = 1 - \text{trace}(\boldsymbol{\chi})$. From equation (52), this is the effective dielectric function, which is the RPA dielectric function for the effective (local field corrected) potential, $v^{\text{eff}}(\mathbf{q})$.

$\hat{D}$ $= \hat{D}(\mathbf{q}, \omega) = D(\mathbf{q}, \omega)/D(\mathbf{q}, 0)$

$D_a$ $= D_a(\mathbf{q}, \omega) = 1 + \Lambda_{aa}(\mathbf{q})\left(\varepsilon_a^0(\mathbf{q}, \omega) - 1\right)$

$D(\lambda)$ (Appendix A). Function defined by $D(\lambda) = |\mathbf{1} + \lambda \mathbf{P}|$

$\hat{\mathbf{E}}_a$ Operator corresponding to the electric field due to charge species $a$

$E_\alpha$ Energy of the microstate $\alpha$

$\dot{\mathcal{E}}_{ab}$ Energy transfer rate, from $a \to b$, between species subsystems $a$, $b$ due to the those subsystems not being in mutual thermal equilibrium.

$\dot{\mathcal{E}}_a$ Net energy transfer rate, to species $a$ in a multicomponent system comprising subsystems $a$, $b$ … due to the those subsystems not being in mutual thermal equilibrium.

$e$ Charge on an electron.

e Euler's constant, or label denoting electron.

$F(q)$ Function defined by (84).

$F_{ab}$ $F_{ab}(\mathbf{q}, \omega)$, function defined by (94).

$\tilde{F}_{ab}$ $= \tilde{F}_{ab}(\mathbf{q}, \omega)$, function defined by (107).

$f_a$     $f_a(\omega) = \omega \coth\left(\dfrac{\omega}{2T_a}\right)$

$\mathfrak{g}$     $= \|g_{ab}\|$, static pair correlation function matrix.

$\mathfrak{g}^0$     $= \|\delta_{ab} g^0_{aa}\|$, exchange pair correlation matrix for non-interacting particles: $g^0_{aa}(\mathbf{q}) = R_a(\mathbf{q}) - 1$.

$\tilde{\mathfrak{g}}$     Intermediate (non-symmetric) pair correlation function defined by (111).

$\mathfrak{g}'$     Boltzmann pair correlation function, $\mathfrak{g}'(\mathbf{q}) = \left\| \dfrac{\tilde{g}_{ab}(\mathbf{q}) - \delta_{ab} g^0_{aa}(\mathbf{q})}{1 + g^0_{bb}(\mathbf{q})} \right\|$, which is equivalent to

$$g'_{ab}(\mathbf{q}) + \delta_{ab} = \dfrac{\tilde{g}_{ab}(\mathbf{q}) + \delta_{ab}}{g^0_{bb}(\mathbf{q}) + 1} = \dfrac{1}{R_b(\mathbf{q})}\left(\tilde{g}_{ab}(\mathbf{q}) + \delta_{ab}\right).$$

$g_{ab}(\mathbf{q}) = \sqrt{n_a n_b} \int e^{-i\mathbf{q}\cdot\mathbf{r}} g_{ab}(\mathbf{r}) d^3\mathbf{r}$, Fourier transform of the two-particle correlation function.

$g^b_{ei}$     Pair correlation function for bound electrons in a single centre problem, defined to be the Fourier transform of the bound electron density normalised according to (151).

$g^f_{ei}$     Pair correlation function for free electrons in a single-centre problem, defined as per (150) ff.

$g^*_{ei}$     Single-centre electron-ion pair correlation function, given by (153),

$g^*_0$     Jellium correlation function, as used in a single-centre atomic calculation..

$\hat{H}$     Hamiltonian operator.

$I_j(x)$     Fermi function defined by (86).

$I, I^0$     Functions defined by (196), (202).

i     $\sqrt{-1}$, or label denoting a positive ion comprising a bare atomic nucleus.

$\mathbf{J}$     Matrix defined by (27).

$J, J^0$     Functions defined by (197),(203).

$\hat{\mathbf{j}}_a$     Charge current operator for species $a$.

$K$     Function defined by (204).

$k$     Debye-Hückel screening parameter, given by $k^2 = \sum_a k_a^2$.

$k_a$     Debye-Hückel screening parameter for species $a$, defined by $k_a^2 = Q_a^2 e^2 n_a / \varepsilon_0 T_a$.

$\mathbf{L}$     Matrix defined by $\mathbf{L} = \mathbf{1} + \dfrac{\boldsymbol{\chi}}{D} = (\mathbf{1} - \boldsymbol{\chi})^{-1}$.

$L, L^0$    Functions defined by (205).

$M_a(\mathbf{q})$ Effective mass of particle species $a$ as defined by (216).

$m_a$    Mass of particle of species $a$.

$N_a$    Number of particles of species $a$ occupying a macroscopic volume $\mathfrak{V}$.

$n_a$    Average density of particles of species $a$, $= \langle N_a \rangle / \mathfrak{V}$.

$n_e^*$    Effective electron density = asymptotic far-field (free) electron density in a single-centre atomic model, in terms of which the degeneracy parameter is given by (88).

$\hat{n}_a$    $= \hat{n}_a(\mathbf{r},t)$, density operator for species $a$.

$\hat{O}$    General operator on the quantum Hilbert space.

$\mathbf{P}$    $= \|P_a\|$, with components given by (19).

$\mathbf{P}$    (Appendix A) General idempotent matrix.

$P_\alpha$    Probability the microstate $\alpha$ within a quantum statistical state.

$p(k) = \left[1 + \exp\left(k^2/2m_e T_e - \eta_e\right)\right]^{-1}$, free-electron Fermi-distribution.

$\mathbf{Q}$    $= \|Q_a\|$ where $Q_a$ is the charge carried by a particle of species $a$, in units of the electronic charge, $|e|$.

$\mathbf{q}$    Momentum/ wavevector reciprocal-space vector.

$R_a$    $= R_a(\mathbf{q}) = 1 + g_{aa}^0(\mathbf{q})$, which is the static structure factor for non-interacting fermi particles as defined by the second of equations (133). This is the same as $R_\mathbf{q}^a$ defined in ref.[28].

$\tilde{R}_a$    $= \tilde{R}_a(\mathbf{q}) = -\dfrac{T_a}{n_a}\Pi_{aa}^0(\mathbf{q},0)$ which definition is equivalent to the first of equations (133). This is the same as $\tilde{R}_\mathbf{q}^a$ defined in ref.[28].

$R_0$    Ion sphere radius.

$\mathbf{r}$    Spatial coordinate vector.

$\mathbf{S}(\mathbf{q},\omega) = \|S_{ab}(\mathbf{q},\omega)\|$, dynamic structure factor matrix.

$\mathbf{S}(\mathbf{q}) = \|S_{ab}(\mathbf{q})\|$, static structure factor matrix.

$\widehat{\mathbf{S}}(\mathbf{q},\omega) = \|\widehat{S}_{ab}(\mathbf{q},\omega)\|$, symmetrised dynamic structure factor matrix, as defined by (33).

$S_{aa}^0(\mathbf{q},\omega)$ Dynamic structure factor for non-interacting particles $a$.

| | |
|---|---|
| $\mathfrak{s}$ | $= \|\mathfrak{s}_{ab}\|$, matrix defined by (114) or (115), whose diagonal elements yield the single component static structure factors., $\mathfrak{s}_{aa}(\mathbf{q}) = S_{aa}(\mathbf{q})$. |
| $T$ | Equilibrium temperature. |
| $T_a$ | Temperature of subsystem comprising particle species $a$. |
| $T_a^{\text{eff}}(\mathbf{q})$ | Effective temperature of particle species $a$ as defined by (216). |
| $T_{ab}^{\text{rel}}(\mathbf{q})$ | Effective relative temperature of two particle species $a$ and $b$, as defined by (217) |
| $t$ | Time. |
| $\hat{V}_a$ | Electrostatic potential due to particles of species $a$. |
| $\mathfrak{V}$ | Volume. |
| $v(\mathbf{q})$ | Standard Coulomb potential, $= e^2/\varepsilon_0 q^2$. |
| $\mathbf{v}(\mathbf{q})$ | $= \|v_{ab}(\mathbf{q})\| = \|Q_a Q_b v(\mathbf{q})\|$, $v_{ab}$ = Coulomb potential acting between particles of species $a$ and $b$. |
| $\mathbf{v}^{\text{eff}}(\mathbf{q})$ | $= \|v_{ab}^{\text{eff}}(\mathbf{q})\| = \|\Lambda_{ab}(\mathbf{q}) v_{ab}(\mathbf{q})\|$ = effective potential, with static Local Field Corrections, as per (48). |
| $\mathbf{X}$ | Matrix defined by (47). |
| $\mathbf{Y}$ | Thermal coupling matrix defined by (45). |
| $Z$ | (Average) nuclear or ionic charge, $= n_e/n_i$. |
| $Z_b$ | Average number of bound electrons per ion. |
| $Z^*$ | $= n_e^*/n_i$. |
| $\alpha$ | labels a microstate of a quantum many-body system. |
| $\Delta(q)$ | Plasma quantum correction as defined by (82) and given by (83). |
| $\Delta_{ab}$ | Antisymmetric matrix of functions defined by $\Delta_{ab}(\omega) = f_a(\omega) - f_b(\omega)$ |
| $\Delta_{abc}$ | $\Delta_{abc}(\mathbf{q},\omega)$, antisymmetric matrix of functions defined by (100) |
| $\delta(x)$ | Dirac delta function. |
| $\delta_{ab}$ | Kronecker delta |
| $\delta \hat{n}_a$ | Density fluctuation operator $= \hat{n}_a - n_a$. |
| $\varepsilon(\mathbf{q},\omega)$ | Dielectric function for Coulomb system. (See also $D(\mathbf{q},\omega)$). |

$\varepsilon_a^0(\mathbf{q},\omega)$   RPA dielectric function for species $a$.

$\varepsilon_0$   Permittivity of free space.

$\eta_e$   Electron degeneracy parameter.

$\mathbf{K}$   $= \|K_{ab}\|$, Green function matrix giving the density response to an external field.

$\mathbf{K}^\dagger$   $= \|K_{ba}^*\|$ = Hermitian Conjugate of $\mathbf{K}$

$\mathbf{K}'$   $= \tfrac{1}{2}(\mathbf{K} + \mathbf{K}^\dagger)$ = Hermitian part of $\mathbf{K}$

$\mathbf{K}''$   $= -\tfrac{1}{2}i(\mathbf{K} - \mathbf{K}^\dagger)$ = Antihermitian part of $\mathbf{K}$ (where $\mathbf{K}''$ is Hermitian).

$K_{aa}^0$   $K_{aa}^0(\mathbf{q},\omega)$, RPA response function, $= \Pi_{aa}^0(\mathbf{q},\omega)/(1 - v_{aa}(\mathbf{q})\Pi_{aa}^0(\mathbf{q},\omega))$

$\tilde{K}_a$   $\tilde{K}_a(\mathbf{q},\omega)$, function defined by (69), $= \Pi_{aa}^0(\mathbf{q},\omega)/(1 - v_{aa}^{\text{eff}}(\mathbf{q})\Pi_{aa}^0(\mathbf{q},\omega))$.

$\hat{K}_{ab}$   $\hat{K}_{ab}(\mathbf{q},\omega) = \dfrac{K_{ab}(\mathbf{q},\omega)}{\Pi_{bb}^0(\mathbf{q},0)}$

$\kappa_a^T$   Isothermal bulk modulus associated with species $a$, where $\kappa_a^T = \dfrac{\langle N_a \rangle}{\langle \Delta N_a^2 \rangle} n_a T_a$.

$\Lambda$   Argument of Coulomb Logarithm.

$\Lambda_{ab}$   $= \Lambda_{ab}(\mathbf{q})$, Local field correction.

$\lambda$   (Appendix A) General scalar parameter.

$\xi_{\text{ei}}$   $= \xi_{\text{ie}}$, electron-ion exchange coefficient, defined by (78).

$\Pi$   $= \|\Pi_{ab}(\mathbf{q},\omega)\|$, Susceptibility, or polarization part of the response matrix $\mathbf{K}$.

$\Pi^0$   $= \|\Pi_{ab}^0(\mathbf{q},\omega)\| = \|\delta_{ab}\Pi_{aa}^0(\mathbf{q},\omega)\|$, RPA susceptibility, which is the polarization part of the RPA response matrix $\mathbf{K}^0$.

$\Pi_{aa}^0$   $= \Pi_{aa}^0(\mathbf{q},\omega)$, Lindhard response function for non-interacting particles. ($\Pi_{aa}^0(\mathbf{q},0) = -n_a \tilde{R}_a(\mathbf{q})/T_a$.)

$\hat{\Pi}_{aa}^0$   $\hat{\Pi}_{aa}^0(\mathbf{q},\omega) = \Pi_{aa}^0(\mathbf{q},\omega)/\Pi_{aa}^0(\mathbf{q},0)$

$\boldsymbol{\rho}$   $= \|\rho_{ab}\|$, where $\rho_{ab}(\mathbf{r})$ is the pair distribution function for species $a$ with respect to species $b$, $= n_a(1 + g_{ab}(\mathbf{r}))$

$\hat{\rho}$   Statistical operator.

$\rho_0^*$   Electron density in jellium.

$\rho_c$     Charge distribution surrounding a central ion.

$\rho_{ei}^*$     Electron distribution surrounding a central ion in a single-centre atomic calculation (jellium model).

$\Sigma(\mathbf{r})$     Function defined by (142).

$\Sigma_0(\mathbf{r})$     Function defined by (145).

$\boldsymbol{\sigma}$     $= \|\sigma_{ab}\|$, matrix defined by (116).

$\sigma_{ab}(\mathbf{q},\omega)$ (Appendix B) Spectral density.

$\boldsymbol{\Phi}$     $= \boldsymbol{\chi}/(1-D)$

$\phi_{ab}$     $\phi_{ab}(\mathbf{q},\omega)$, function defined by (97).

$\boldsymbol{\chi}$     $= \|\chi_{ab}(\mathbf{q},\omega)\|$, Polarizability matrix.

$\boldsymbol{\chi}_0$     $= \|\chi_{ab}(\mathbf{q},0)\|$, Static polarizability.

$\Omega_\mathbf{q}$     Frequency of ion plasma mode with wavevector $\mathbf{q}$.

$\Omega_i$     Ion plasma frequency.

$\omega$     Frequency.

This page is intentionally left blank